\documentclass[10pt]{article}

\usepackage{graphicx}
\usepackage{indentfirst,csquotes}
\usepackage{amssymb,amsthm,amsmath}
\usepackage{xcolor,paralist,hyperref,titlesec,fancyhdr}
\hypersetup{
  colorlinks=true,
  citecolor=blue,
  linkcolor=blue,
  urlcolor=blue
}

\usepackage{lipsum}
\usepackage{siunitx}    
\usepackage{natbib}    
\usepackage{booktabs}   
\usepackage{float}
\usepackage{braket} 
\usepackage{multirow} 
\usepackage{makecell}
\newenvironment{acknowledgments}{\section*{Acknowledgments}}{}

\titleformat{\section}{\normalfont\large\bfseries}{\thesection}{1em}{}
\titlespacing*{\section}{0pt}{1ex}{1ex}

\hypersetup{ colorlinks=true, linkcolor=red, filecolor=red, urlcolor=red }

\begin{document}

\title{Free-space and Satellite-Based Quantum Communication: Principles, Implementations, and Challenges}
\author{
  Georgi Gary Rozenman$^{1}$ \and
  Alona Maslennikov$^{2}$ \and
  Sara P.~Gandelman$^{3,4}$ \and
  Yuval Reches$^{3}$ \and
  Sahar Delfan$^{5}$ \and
  Neel Kanth Kundu$^{6,7}$ \and
  Leyi Zhang$^{8}$ \and
  Ruiqi Liu$^{8}$
}

\date{%
  $^{1}$ Department of Mathematics, Massachusetts Institute of Technology,  
  Cambridge, Massachusetts 02139, USA\\
  $^{2}$ Department of Chemistry, Boston University,  
  Boston, Massachusetts 02215, USA\\
  $^{3}$ The Raymond and Beverly Sackler School of Physics and Astronomy,  
  Tel Aviv University, Tel Aviv 69978, Israel\\
  $^{4}$ School of Electrical Engineering, Iby and Aladar Fleischman Faculty of Engineering,  
  Tel Aviv University, Tel Aviv 69978, Israel\\
  $^{5}$ Institute for Quantum Science and Engineering, Department of Physics and Astronomy,  
  Texas A\&M University, College Station, Texas 77843, USA\\
  $^{6}$ Centre for Applied Research in Electronics (CARE) and Bharti School of Telecommunication Technology and Management, 
  Indian Institute of Technology Delhi, New Delhi 110016, India\\
  $^{7}$ Department of Electrical and Electronic Engineering, University of Melbourne, Melbourne, VIC, Australia.\\
  $^{8}$ Wireless and Computing Research Institute, ZTE Corporation,  
  Beijing 100029, China\\[1ex]
  \texttt{garyrozenman@protonmail.com}\\[2ex]
  \today
}

\maketitle

\begin{abstract}
\quad
Satellite-based quantum communications represent a critical advancement in the pursuit of secure, global-scale quantum networks. Leveraging the principles of quantum mechanics, these systems offer unparalleled security through Quantum Key Distribution (QKD) and other quantum communication protocols. This review provides a comprehensive overview of the current state of satellite-based quantum communications, focusing on the evolution from terrestrial to space-based systems. We explore the distinct advantages and challenges of discrete-variable (DV) and continuous-variable (CV) quantum communication technologies in the context of satellite deployments. The paper also discusses key milestones such as the successful implementation of quantum communication via the Micius satellite and outlines the primary challenges, including atmospheric turbulence and the development of quantum repeaters, that must be addressed to achieve a global quantum internet. This review aims to consolidate recent advancements in the field, providing insights and perspectives on the future directions and potential innovations that will drive the continued evolution of satellite-based quantum communications.\end{abstract} 

\bigskip

\begin{quotation}
In an era where information security is critical, quantum communication has emerged as a revolutionary technology that offers unprecedented protection for data transmission. Unlike classical communication methods, which are vulnerable to interception and cyberattacks, quantum communication utilizes the fundamental laws of quantum mechanics to ensure theoretically unbreakable security. The development of satellite-based quantum communication represents a major breakthrough, extending secure links beyond the limits of terrestrial infrastructure and enabling the foundation of a global quantum network. This paper examines the current state of satellite-based quantum communications, highlighting significant advancements in both discrete-variable and continuous-variable technologies while addressing the challenges that remain. As global investment in quantum satellites accelerates, the realization of a secure and interconnected quantum internet is approaching. This positions satellite-based quantum communication at the forefront of next-generation secure technologies. \end{quotation}

\section{Overview of Quantum Communications}

Since the early days of civilization, human progress has been closely related to the advancement in communication. From smoke signals and carrier pigeons to the invention of the telegraph, radio, and the internet, each technological innovation has fundamentally transformed the way societies interact and share information. In today’s hyper-connected world, communication networks are the foundation of global infrastructure, allowing seamless interactions over vast distances \cite{huurdeman2003worldwide}. However, with our growing reliance on digital communication comes an urgent need for secure and tamper-proof information exchange \cite{merkle1978secure,abu2017obstacles,vo2021economic}. While classical cryptographic methods have proven to be effective, they remain vulnerable to increasing computational power and evolving cyber threats \cite{tambe2023household,sarsam2023cybersecurity}. This growing threat has driven a paradigm shift toward quantum communication, a revolutionary technology that leverages the laws of quantum mechanics to ensure fundamentally secure information transfer. Among the most promising developments is satellite-based quantum communication, which extends the capabilities of terrestrial quantum networks and lays the groundwork for a global quantum-secure infrastructure \cite{dai2020towards,yavuz2022distributed,lewis2022secure, del2024cybersecurity,zornetta2024quantum}.

As societies continue to benefit from the ubiquitous connectivity enabled by advanced communication systems \cite{uzoka2024role}, security has become a critical concern in all types of networks \cite{frustaci2017evaluating}. Both private users and vertical industries demand a high level of security \cite{monshizadeh2021security}, which can be supported by quantum communication \cite{wang2022quantum,turnip2025towards}.

Quantum communication represents a groundbreaking shift in the way secure information transfer is achieved \cite{cariolaro2015quantum,sidhu2021advances}, leveraging the principles of quantum mechanics to provide unprecedented levels of security \cite{shafique2024hybrid,bhosale2023quantum}. The journey from terrestrial to satellite-based quantum communication systems marks a significant advancement in overcoming the limitations of distance, channel loss, and eavesdropping vulnerabilities that conventional communication systems face \cite{sidhu2021advances,cariolaro2015quantum}. Quantum Key Distribution (QKD), the most prominent application of quantum communication, enables two parties to share a cryptographic key with unconditional security, guaranteed by the laws of quantum physics \cite{turnip2025towards}. However, the implementation of QKD over long distances has been challenging due to the attenuation and decoherence effects in optical fibers and Free-space channels \cite{peelam2024quantum,rozenman2023quantum}. The deployment of quantum communication systems in space via satellites offers a promising solution to these challenges, paving the way for the establishment of a global quantum network \cite{frustaci2017evaluating}.

\begin{figure*}[htbp]
    \centering
    \scalebox{0.4}{\includegraphics{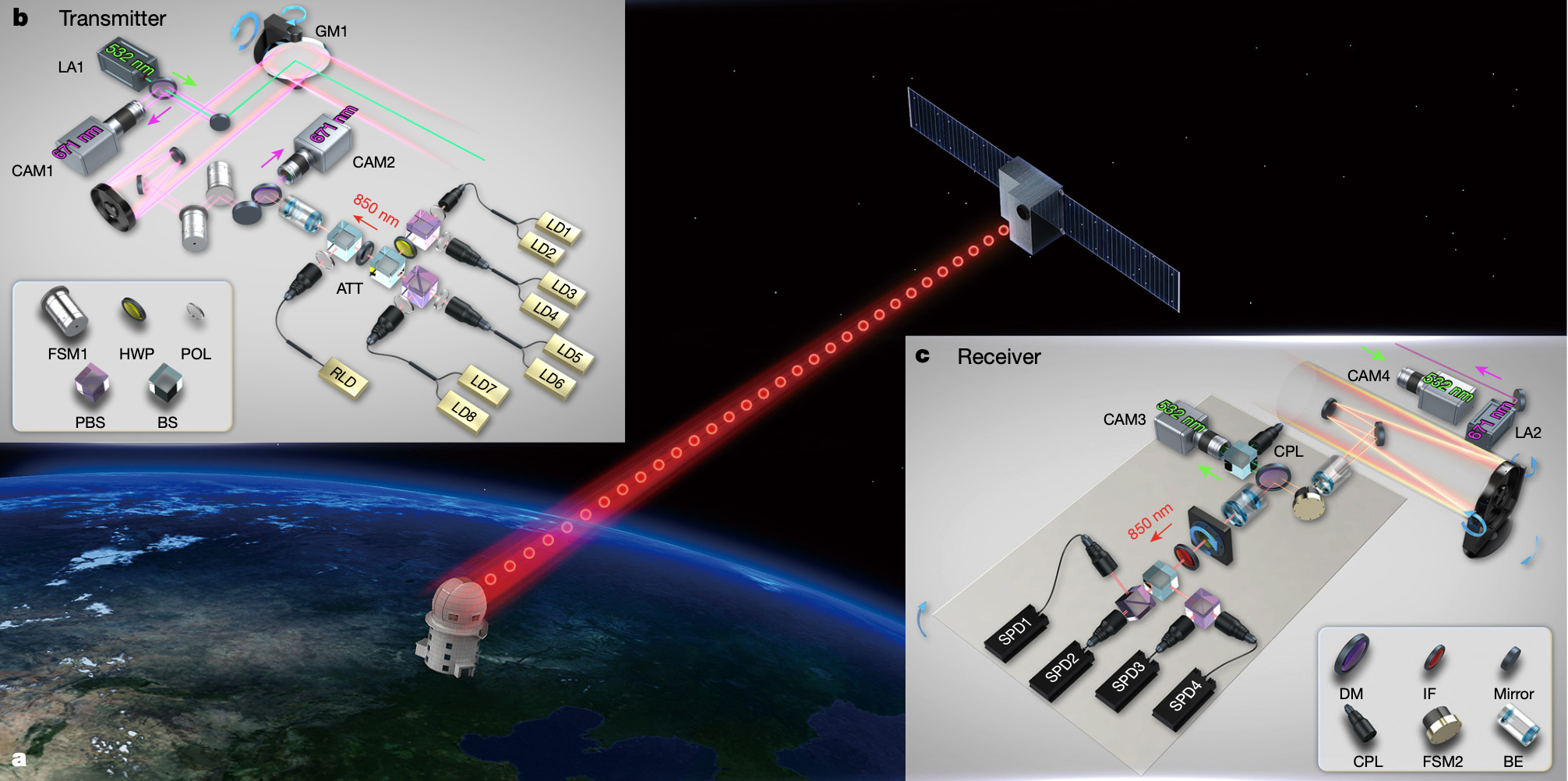}}
     \caption{Experimental setup for satellite-to-ground quantum key distribution (QKD) using the Micius satellite \cite{liao2017satellite}. The Micius satellite, weighing 635 kg, operates in a Sun-synchronous orbit approximately 500 km above Earth and carries three payloads designed for space-based quantum experiments including QKD, Bell tests, and quantum teleportation. The satellite’s QKD transmitter employs eight laser diodes emitting attenuated pulses at around 850 nm, which pass through a BB84 encoding module composed of polarizing beam splitters, a half-wave plate, and a beam splitter. The encoded quantum signals are co-aligned with a 532 nm green laser used for system tracking and time synchronization, then transmitted via a 300-mm-aperture Cassegrain telescope. Beam control is achieved through a two-axis gimbal mirror for coarse tracking and fast steering mirrors for fine tracking, while a low-power 671 nm laser serves as a polarization reference. On the ground, the Xinglong station features a 1,000-mm-aperture telescope that separates the incoming 532 nm tracking laser and 850 nm quantum signals using a dichroic mirror. The tracking beam is monitored by a camera for alignment, while the quantum signals are analyzed by a BB84 decoder consisting of beam splitters and four single-photon detectors. The ground station also sends a 671 nm laser beam back to the satellite for reciprocal tracking. This dual-wavelength synchronization and hybrid tracking system enable precise alignment and polarization compensation, facilitating high-rate QKD over distances up to 1,200 km and demonstrating a significant advancement in space-based quantum communication.}
    \label{Fig_1}
\end{figure*}

The motivation for satellite-based quantum communications arises from the need to overcome the distance limitations inherent in terrestrial quantum networks \cite{NSC}. Quantum communication relies on fundamental phenomena such as entanglement and superposition, which enable secure information transfer, a capability that classical systems cannot match. Although terrestrial QKD systems perform effectively over shorter distances, they experience exponential signal loss in optical fibers and Free-space channels, restricting their operational reach to a few hundred kilometers. Satellite-based systems, on the other hand, can transmit quantum signals between satellites and ground stations, allowing secure links over significantly greater distances. By bypassing the limitations of Earth's curvature and reducing exposure to atmospheric attenuation, these systems offer a viable solution for establishing global-scale quantum communication networks.

The launch of China's Micius satellite in 2016 marked a major milestone in the advancement of quantum communication, demonstrating the feasibility of satellite-based QKD on a global scale. Micius successfully enabled QKD over thousands of kilometers, validating the robustness of quantum communication protocols under real space conditions. By enabling secure links between distant ground stations, satellite platforms like Micius pave the way for a future global quantum internet, one in which secure information exchange is possible between any two locations on Earth. 

\section{Key Protocols in Free-space Quantum Key Distribution}

Quantum Key Distribution (QKD) is a secure communication technique that leverages the principles of quantum mechanics to generate and distribute cryptographic keys \cite{wolf2021quantum}. Unlike classical encryption methods, QKD offers unconditional security; its robustness is not compromised even by adversaries with unlimited computational resources. This makes QKD an ideal solution for high-security applications in sectors such as finance, government, and defense \cite{lovic2020quantum}. Several QKD protocols have been developed over the years, each with distinct features and advantages. Among the most widely used are the BB84, B92, and Ekert protocols. Although they differ in the way they encode and transmit key information, they all rely on the same fundamental principles of quantum mechanics, such as the Heisenberg uncertainty principle and the no-cloning theorem \cite{wolf2007quantum,wolf2021quantum}

Regardless of the protocol, the objective of all QKD systems remains the same: to allow two parties to securely generate a shared key that can be used for symmetric encryption. This process ensures that any eavesdropping attempt introduces detectable disturbances, preserving the security of the key exchange \cite{jha2019survey}.

\subsection{The BB84 protocol}

\begin{figure*}[htbp]
		\scalebox{0.3}{\includegraphics{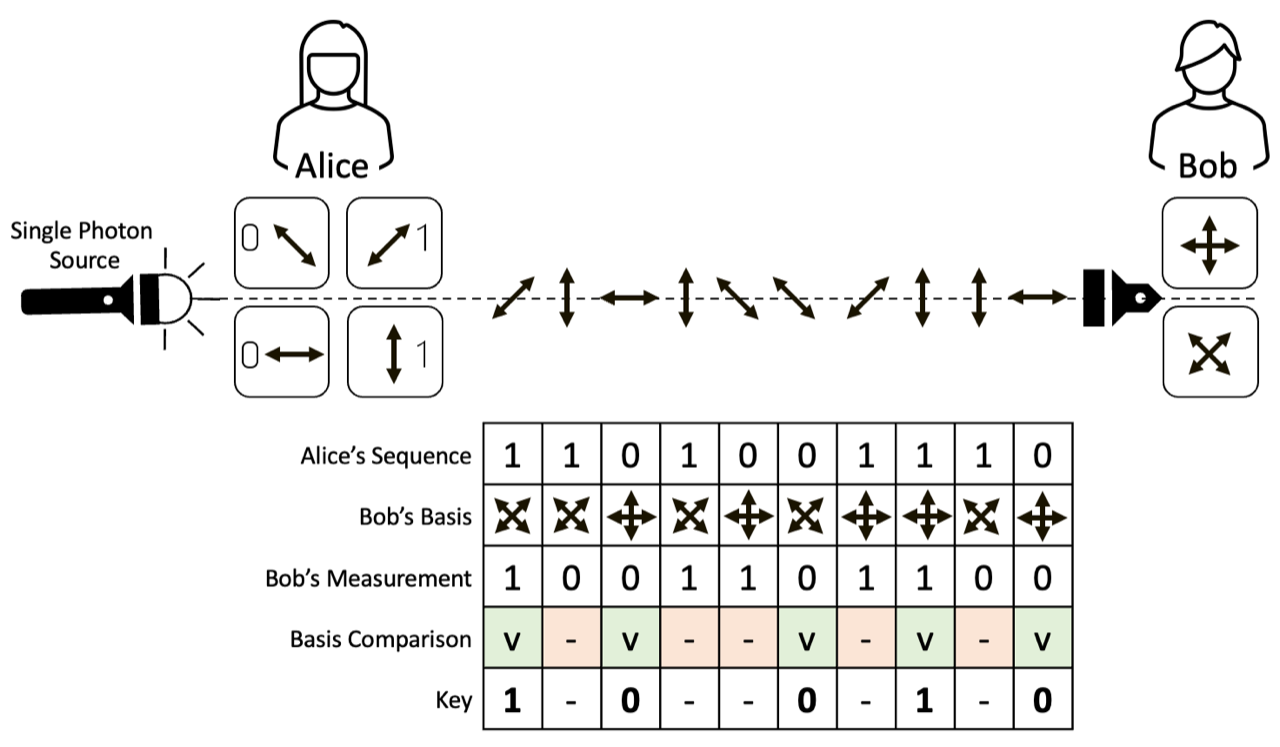}}
		\caption{
  Schematic diagram of a QKD system based on the BB84 protocol with a Free-space communication channel.  Two parties, typically named Alice and Bob, wish to communicate securely over a public channel. Alice transmits a series of individual photons to Bob, each with a random polarization state (horizontal, vertical, diagonal, or anti-diagonal). Bob then measures the polarization of each photon in a randomly chosen basis (horizontal, vertical or diagonal, anti-diagonal) and records the result. Afterwards, Alice and Bob compare a subset of their measurements to detect any eavesdropping attempts.}
	    \label{Fig_2}
\end{figure*}

The BB84 protocol, introduced by Charles Bennett and Gilles Brassard in 1984 \cite{bennett1984proceedings}, was the first practical quantum key distribution protocol and remains foundational to the field. As illustrated in Fig. \ref{Fig_2}, the security of the protocol is based on two key assumptions: (1) Information gain is only possible at the cost of disturbing the quantum state when non-orthogonal states are used, an effect rooted in the No-Cloning Theorem, and (2) the presence of an authenticated public classical communication channel between the sender and receiver.

In the BB84 protocol, Alice seeks to securely transmit a private key to Bob. She begins with two $n$-bit random strings, $a = (a_1, \dots, a_n)$ and $b = (b_1, \dots, b_n)$. The string $a$ determines the raw key bits, while $b$ specifies the basis used for encoding each qubit. For each index $i$, Alice encodes $a_i$ into a qubit as follows: if $b_i = 0$, she uses the computational basis $\{\ket{0}, \ket{1}\}$, preparing $\ket{0}$ if $a_i=0$ and $\ket{1}$ if $a_i=1$; if $b_i = 1$, she uses the Hadamard basis $\{\ket{+}, \ket{-}\}$, where $\ket{\pm} = (\ket{0} \pm \ket{1})/\sqrt{2}$, preparing $\ket{+}$ if $a_i=0$ and $\ket{-}$ if $a_i=1$.
\cite{bennett1984proceedings}.

The four qubit states below are used to describe the protocol states \cite{Bloom2021,bruss2000quantum},

\begin{equation} \label{eq1}
\begin{split}
& \left|\psi_{00}\right\rangle=|0\rangle \\
& \left|\psi_{10}\right\rangle=|1\rangle \\
& \left|\psi_{01}\right\rangle=|+\rangle=\frac{1}{\sqrt{2}}|0\rangle+\frac{1}{\sqrt{2}}|1\rangle, \\
& \left|\psi_{11}\right\rangle=|-\rangle=\frac{1}{\sqrt{2}}|0\rangle-\frac{1}{\sqrt{2}}|1\rangle .
\end{split}
\end{equation}

The bit $b_{i}$ determines which basis $a_{i}$ is encoded in (either in the computational basis or the Hadamard basis). It is impossible to distinguish each qubit with certainty without knowing $b$ because the qubits are currently in states that are not mutually orthogonal.

Through the public and verified quantum channel $\varepsilon$, Alice sends Bob state $\psi$. Bob receives a quantum state described by $\varepsilon(\rho)=\varepsilon(\ket{\psi}\!\bra{\psi})$, where $\varepsilon$ denotes the combined effects of channel noise and potential eavesdropping, whom we'll refer to as Eve. Both Bob and Eve have their own states after receiving the string of qubits \cite{bruss2000quantum}. However, since only Alice is aware of this, it is essentially impossible for Bob or Eve to tell the states of the qubits apart. Additionally, the no-cloning theorem tells us that Eve cannot be in possession of a copy of the qubits sent to Bob after Bob has received them unless she has taken measurements. But if she chooses the incorrect basis for her measurements, she runs the risk of disturbing a specific qubit with probability 1/2.

In summary, as shown in Fig. \ref{Fig_2},

\begin{enumerate}
    \item Alice selects two classical bit strings $a = (a_1, \dots, a_{4n})$ and $b = (b_1, \dots, b_{4n})$. The string $a$ encodes the raw key bits, while $b$ specifies the encoding bases.
    \item For each index $i$, Alice encodes $a_i$ into a qubit: if $b_i = 0$ she uses the computational basis $\{\ket{0},\ket{1}\}$, and if $b_i = 1$ she uses the Hadamard basis $\{\ket{+},\ket{-}\}$. She thus prepares a block of $4n$ qubits and sends them to Bob over the quantum channel.
    \item Bob receives the $4n$ qubits and, for each $i$, chooses a measurement basis at random: computational ($0$) or Hadamard ($1$). His outcomes define bit strings $a'$ and $b'$, where $b'$ records his basis choices.
    \item Alice publicly reveals her basis string $b$. Bob compares it with his own $b'$, and they keep only the positions where $b_i = b'_i$. This “sifting” step leaves, on average, $2n$ shared bits.
    \item To estimate the error rate (and possible eavesdropping), Alice and Bob publicly compare a random subset of $n$ of these bits. If the error rate is too high, they abort; otherwise, they proceed.
    \item Finally, Alice and Bob apply classical error correction and privacy amplification protocols to the remaining $\sim n$ bits, distilling a shorter but secure shared secret key of length $m$ bits.
\end{enumerate}

\subsection{The E91 Protocol}

Entanglement can be used effectively to establish a secret key between two parties. In particular, entangled photon pairs provide inherent security advantages. For example, photon-number-splitting (PNS) attacks \cite{lib2022processing}  are significantly less successful in entanglement-based protocols, as it is highly unlikely to simultaneously generate multiple entangled photon pairs \cite{ekert1991quantum}. 

In such protocols, a source generates and distributes entangled states between Alice and Bob. This source can be physically located anywhere and operated by any entity, including a potentially malicious one. For example, the entangled pairs may be prepared locally by Alice before the protocol is executed, or distributed to Alice and Bob by a third party, commonly known as Charlie \cite{gordon2010quantum}. 

In the most adversarial scenario, the source of entangled states may be entirely controlled by Eve. Consequently, it is treated as untrusted, and security analyses often assume a worst-case situation in which Eve has full access to the state preparation process. Alice and Bob, however, share an authenticated classical communication channel over which Eve can eavesdrop but cannot alter messages. Since the entangled states are distributed prior to the protocol’s execution and no direct quantum channel between Alice and Bob is required, entanglement-based QKD protocols often simplify the overall security analysis.

According to the protocol described in \cite{yamamoto2016principles}, Alice and Bob have access to a source that distributes maximally entangled pairs of qubits in the following state:

\begin{equation}
    \mid \Psi^{-}\rangle_{AB} = \frac{1}{\sqrt{2}} (\mid 01 \rangle _{AB} - \mid 10 \rangle _{AB} )  
\end{equation}

A sequence of $\mid \Psi^{-}\rangle_{AB}$ states is distributed, with the first qubit of each pair assigned to Alice and the second to Bob. For each entangled pair, Alice and Bob randomly choose their measurement settings from predefined sets ${Ai}$ and ${Bi}$, respectively. Then, they publicly announce their measurement bases. In cases where the chosen directions match, i.e., $(A1, B1)$ and $(A3, B3)$, the results form the sifted key. In contrast, results corresponding to mismatched bases between Alice and Bob, i.e., $(A1, B3)$, $(A1, B2)$, $(A2, B3)$, and $(A2, B2)$ are used to evaluate a CHSH inequality, given by \cite{arbel2025optical}:

\begin{equation}
    S=\mid \langle A_{1} B_{3} \rangle + \langle A_{1} B_{2} \rangle  + \langle A_{2} B_{3} \rangle + \langle A_{2} B_{2} \rangle \mid 
\end{equation}

In ideal quantum systems, the CHSH parameter reaches the theoretical maximum of  $S=2 \sqrt{2}$ . However, in realistic implementations, $S>2$ suffices to demonstrate quantum correlations and the security of the distributed key.

\subsection{The B92 Protocol}

The B92 protocol, introduced by Bennett in 1992~\cite{bennett1992quantum}, is a simplified QKD protocol that uses only two non-orthogonal quantum states. Unlike the BB84 protocol that employs four states, B92 relies on the fundamental principles of the no-cloning theorem and measurement-induced disturbance to ensure security against eavesdropping. In this scheme, Alice encodes classical bits using the following non-orthogonal polarization states:

\begin{equation}
\label{eq:b92_states}
\begin{aligned}
|\Psi_{00}\rangle &= |0^\circ\rangle = |H\rangle \\
|\Psi_{01}\rangle &= |45^\circ\rangle = \frac{1}{\sqrt{2}}(|H\rangle + |V\rangle)
\end{aligned}
\end{equation}

Upon receiving each quantum state, Bob randomly selects a measurement basis - either the rectilinear basis $\{0^\circ, 90^\circ\}$ or the diagonal basis $\{45^\circ, -45^\circ\}$. A measurement yields a conclusive result only if Bob's chosen projector is orthogonal to the state \emph{not} sent by Alice. These conclusive outcomes, which occur with a probability of $1 - |\langle \Psi_{00} | \Psi_{01} \rangle|^2$, form the raw key. 

The protocol’s security is fundamentally rooted in the fact that non-orthogonal quantum states cannot be perfectly distinguished. Any eavesdropper attempting to intercept and measure the quantum state inevitably introduces detectable disturbances, making the B92 protocol a viable and secure method for QKD despite its minimalist design \cite{gandelman2025hands}.

\subsection{The Six-State BB84 Protocol}

The six-state protocol is a natural extension of the BB84 protocol, offering enhanced security by employing three mutually unbiased bases instead of only two~\cite{bruss1998optimal}. In addition to the computational ($Z$) and diagonal ($X$) bases used in BB84, the six-state protocol incorporates the eigenstates of the Pauli $Y$ operator, which correspond to right- and left-handed circular polarization states:

\begin{equation}
\left|\psi_{y^{+}}\right\rangle = \frac{1}{\sqrt{2}}
\begin{pmatrix} 1 \\ i \end{pmatrix}, \qquad
\left|\psi_{y^{-}}\right\rangle = \frac{1}{\sqrt{2}}
\begin{pmatrix} 1 \\ -i \end{pmatrix}.
\end{equation}

Physically, $\left|\psi_{y^{+}}\right\rangle$ represents right-handed (clockwise) circular polarization, in which the electric field vector rotates in time with the phase of the vertical component leading the horizontal by $\pi/2$. Conversely, $\left|\psi_{y^{-}}\right\rangle$ corresponds to left-handed (counterclockwise) circular polarization, where the vertical component lags by $\pi/2$. These states are orthogonal and normalized, as can be verified directly:

\begin{equation}
\braket{\psi_{y^{+}} | \psi_{y^{-}}}
= \frac{1}{2} \left( 1 \times 1 + i \times (-i) \right)
= \frac{1}{2} (1 - 1) = 0.
\end{equation}

Hence, the two circular polarization states form a complete, orthonormal basis for the $Y$-eigenstates of a qubit, just as horizontal/vertical and diagonal/antidiagonal polarizations do for the $Z$ and $X$ bases, respectively. Together, the six states of the $X$, $Y$, and $Z$ bases span the entire Bloch sphere, providing a complete sampling of qubit state space—unlike the four-state BB84 protocol, which is confined to a single great circle.

The protocol operates analogously to BB84: Alice randomly chooses one of the three encoding bases ($X$, $Y$, or $Z$) to prepare each qubit, while Bob independently measures in a randomly chosen basis. Owing to the addition of the third basis, the probability that Alice and Bob select matching bases decreases from $1/2$ to $1/3$, meaning that approximately two-thirds of the raw bits are discarded during the sifting stage~\cite{bechmann1999incoherent}.

Despite the lower sifting efficiency, the six-state protocol provides increased robustness against eavesdropping. Because Eve must now guess among three mutually unbiased bases, any intercept–resend attempt introduces a higher quantum bit error rate (QBER), making her presence more readily detectable. This expanded set of non-commuting measurements thus strengthens security against a broader class of quantum attacks, including coherent attacks~\cite{mckague2010generalized}, while emphasizing the critical role of circular polarization in achieving full Bloch-sphere coverage and maximal security symmetry \cite{amal2022quantum}.

\subsection{The Decoy State Protocol}

The decoy state protocol is one of the most widely used methods in QKD, offering enhanced security against photon number splitting (PNS) attacks \cite{schmitt2007experimental,rosenberg2007long}. Unlike the ideal BB84 protocol, practical QKD systems often rely on weak coherent pulses, which occasionally emit multiple photons. This makes them susceptible to PNS attacks, where an eavesdropper can redirect a photon from a multiphoton pulse without disturbing the quantum state, compromising security and severely limiting the secure transmission and maximum channel length.

To address this, the decoy state protocol introduces additional intensity levels at the transmitter: one signal state and several decoy states \cite{wang2007practical}. By randomly varying the photon number distribution and announcing the intensity level only after transmission, Alice prevents an eavesdropper from selectively targeting multiphoton sources \cite{shapiro2009defeating}.

A successful PNS attack requires the bit error rate (BER) to remain consistent across all intensity levels, which is an impossible outcome due to the varying photon statistics of the decoy states. As a result, legitimate users can detect attempted PNS attacks by analyzing BERs for each intensity level, while also achieving significantly higher secure key rates and longer channel distances \cite{rajendran2024mitigating}.

The decoy state protocol typically proceeds in three stages \cite{ma2005practical}. 
1. Alice sends a sequence of photon pulses to Bob, randomly selecting between signal, decoy, and vacuum states. The signal states carry the actual quantum key information, while the decoy states serve to monitor the channel, and the vacuum states help estimate background noise and dark counts at the detector.

2. Bob randomly chooses a measurement basis (typically rectilinear or diagonal) for each incoming photon and records the results, including which basis was used.

3. Alice publicly announces the intensity level (signal, decoy, or vacuum) used for each pulse. The two parties reject inconclusive events and use the remaining data to estimate error rates. Only the signal state measurements with matching bases are used for final key generation. If the decoy and vacuum states show unexpected error statistics, the protocol indicates the possibility of eavesdropping.

Implementation of the decoy state protocol significantly enhances the security of QKD systems that use imperfect photon sources, making them viable for real-world applications.

\subsection{The Decoy‐State BB84 with Polarization‐Only Qubits Using Amplitude Modulation}
\label{sec:decoyBB84}

Polarization‐encoded BB84 systems employing weak coherent pulses (WCPs) must mitigate photon-number-splitting (PNS) attacks. Incorporating \emph{decoy} pulses with varied mean photon numbers exposes eavesdroppers through inconsistencies in the single-photon yield \( Y_1 \) and error rate \( e_1 \).
All-polarization implementations are attractive for Free-space, fiber, and space links, as they require no interferometric stability. Recent advances in integrated amplitude modulators now enable three-intensity (``vacuum + weak + signal'') ensembles at multi-hundred-MHz rates while maintaining spectral indistinguishability and closing side channels~\cite{wolf2021quantum}.

Logical bits are encoded in polarization states
\[
\{|H\rangle,|V\rangle\}\;(Z\text{ basis}), \qquad
\{|D\rangle,|A\rangle\}\;(X\text{ basis}),
\]
generated from a single gain-switched laser. A polarization modulator sets the basis and bit, and an ultrafast amplitude modulator controls the mean photon number \( \mu\!\in\!\{\mu_{\rm sig},\mu_{\rm dec},0\} \). Since the intensity is a property of the light pulse and not part of the quantum information encoded in the polarization state, intensity modulation does not disturb the qubit, provided the modulator is polarization-independent \cite{dhingra2023high}.

A LiNbO\textsubscript{3} phase modulator embedded in a Sagnac interferometer converts phase modulation into intensity modulation, enabling per-pulse attenuation at rates exceeding \(600~\mathrm{MHz}\) with an extinction ratio greater than \(30~\mathrm{dB}\). The use of a single laser source ensures identical spectral and temporal properties for all emitted pulses, thereby eliminating ``which-laser'' side channels and suppressing patterning effects that can otherwise compromise security.

In a three-intensity decoy-state implementation, Alice randomly prepares optical pulses with different mean photon numbers. The signal state, characterized by a mean photon number \(\mu_{\text{sig}}\), is used for key generation, while the decoy state, with mean photon number \(\mu_{\text{dec}}\), is used exclusively for channel parameter estimation. A third class of pulses corresponds to the vacuum state with \(\mu_{\text{vac}} = 0\), which enables direct estimation of background and detector dark-count contributions. Typical operating parameters are
\[
\mu_{\text{sig}} \simeq 0.5, \qquad
\mu_{\text{dec}} \simeq 0.1, \qquad
\mu_{\text{vac}} = 0,
\]
with the respective preparation probabilities
\[
p_{\text{sig}} \approx 0.8, \qquad
p_{\text{dec}} \approx 0.1, \qquad
p_{\text{vac}} \approx 0.1.
\]
This choice of intensities and probabilities balances secure key-rate efficiency against multiphoton emission probability, allowing tight bounds on the single-photon yield and error rate while maintaining high throughput in practical Free-space and satellite-based QKD systems~\cite{ma2005practical}.

For phase-randomized WCPs, the photon number follows a Poisson distribution. Consequently, the overall gain \( Q_\mu \) and error gain \( Q_\mu E_\mu \) are given by:
\[
Q_\mu = \sum_{n=0}^{\infty} e^{-\mu}\frac{\mu^n}{n!}Y_n, \qquad
Q_\mu E_\mu = \sum_{n=0}^{\infty} e^{-\mu}\frac{\mu^n}{n!}Y_n e_n.
\]
With two non-zero intensities, linear programming bounds \( Y_1 \) and \( e_1 \) tightly for \( \mu_{\text{dec}}\!\ll\!\mu_{\text{sig}} \). Finite-size corrections typically scale as \( O\!\left(\sqrt{\ln(1/\varepsilon)/N}\right) \).


\subsection{The Coherent One Way Protocol}

The Coherent One-Way (COW) protocol utilizes weak coherent light pulses distributed across time-bin pairs to encode key bits. In this scheme, Alice encodes the bit “0” by sending a pulse in the first time bin and vacuum in the second, while the bit “1” is encoded as vacuum followed by a pulse. To detect eavesdropping and monitor channel integrity, Alice randomly interleaves decoy sequences, typically two consecutive pulses (pulse–pulse) or two consecutive vacuums (vacuum–vacuum)~\cite{roberts2017modulator, chau2018decoy, stucki2005fast}. This approach combines time-bin encoding with coherence monitoring to ensure secure communication over long distances.


Alice's setup begins with a continuous-wave (CW) laser that provides a stable phase reference, essential for monitoring coherence across pulses. The beam is modulated by an intensity modulator that carves it into discrete time-bin pulses according to the desired bit pattern or decoy sequence. Then, an optical attenuator reduces the pulses to weak coherent states (WCP) with a typical photon number $\mu \sim 0.1$--$0.2$. This ensures that most pulses contain either zero or one photon, enabling quantum behavior while using readily available laser sources \cite{jain2020design}.

Bob's setup uses a passive beam splitter (e.g., 90/10) to divide the incoming signal into two separate paths. The majority of the signal (data line) is used for key generation while the remainder (monitoring line) is used for coherence verification. On the data line, a time-resolved single-photon detector (SPD) records the arrival time of photons. Detection in the early time-bin corresponds to the bit ``0,'' while detection in the late time-bin corresponds to the bit ``1.'' On the monitoring line, an unbalanced Mach--Zehnder interferometer (MZI) introduces a delay equal to one time-bin, allowing interference between adjacent pulses. The interferometer has two output ports, each connected to detectors $D_c$ (constructive interference) and $D_d$ (destructive interference). The relative counts at these detectors allow Bob to compute the interference visibility \cite{dadahkhani2025experimental}:
    \begin{equation}
        V = \frac{P(D_c) - P(D_d)}{P(D_c) + P(D_d)},
    \end{equation}
    where $P(D_c)$ and $P(D_d)$ are the observed detection probabilities at each output. High visibility values indicate maintained coherence, while deviations suggest channel noise or eavesdropping.

Bob records all detection events with precise time stamps. After the quantum transmission, he announces the time bins where detection events occurred. Alice then publicly reveals which time slots corresponded to decoy states and which to data, allowing Bob to sift the raw key, estimate the quantum bit error rate (QBER), and evaluate channel coherence \cite{das2024evaluating}.


Let $|\alpha\rangle$ denote a coherent state and $|0\rangle$ the vacuum. Alice's quantum states per time-bin pair are \cite{sandfuchs2025security}:
\begin{align}
    |\phi_0\rangle &= |\alpha\rangle \otimes |0\rangle, \\
    |\phi_1\rangle &= |0\rangle \otimes |\alpha\rangle, \\
    |\phi_D\rangle &= |\alpha\rangle \otimes |\alpha\rangle.
\end{align}

In the monitoring interferometer, adjacent coherent pulses interfere.

The COW protocol provides inherent robustness against PNS attacks. Since an eavesdropper cannot non-invasively measure the time-bin occupancy of individual pulses without destroying their phase coherence, any eavesdropping attempt introduces a measurable drop in interference visibility. This makes visibility monitoring an effective countermeasure against eavesdropping ~\cite{lavie2022improved}.

\subsection{HD-QKD with Qubit-Like States (Fourier-Qubits) Protocol}

High-dimensional quantum key distribution (HD-QKD) represents a significant advancement in quantum communication by encoding information in a larger Hilbert space than the conventional two-dimensional systems used in protocols like BB84 \cite{scarfe2025fast}. This approach leverages degrees of freedom such as orbital angular momentum (OAM) \cite{willner2015optical}, time-bin encoding, and spatial modes, offering improved information capacity per photon, enhanced noise tolerance, and greater security.

Traditional QKD protocols encode one bit per photon using two-level quantum systems (qubits). In contrast, HD-QKD uses qudits---$d$-level quantum systems---to encode more than one bit per photon, thereby improving the key rate and potentially the resistance to eavesdropping.

A recent advance in high-dimensional quantum key distribution (HD-QKD) introduces a protocol that combines the advantages of large Hilbert spaces with the experimental simplicity of qubit-based schemes. In this approach, Scarfe \emph{et al.}\cite{scarfe2025fast} proposed a high-dimensional BB84-like protocol employing so-called \emph{Fourier-qubits} (F-qubits), which are qubit-like superpositions embedded within a $d$-dimensional computational basis \cite{scarfe2025fast}.

Unlike conventional HD-QKD protocols that rely on mutually unbiased bases (MUBs) constructed from balanced superpositions of all $d$ basis states, the F-qubit protocol replaces the Fourier basis with states that are superpositions of only \emph{two} computational basis elements. An F-qubit takes the form
\begin{equation}
\ket{\phi^{(m)}_{jk}} = \frac{1}{\sqrt{2}}\left( \ket{j} + \omega_d^{\,m}\ket{k} \right),
\end{equation}
where $\omega_d = e^{2\pi i/d}$, $j<k$, and $m\in\{0,\ldots,d-1\}$. Although these states are not mutually unbiased with respect to the computational basis, they retain sufficient phase sensitivity to bound Eve’s information via an effective phase-error estimation\cite{zahidy2024practical}.

The protocol proceeds analogously to BB84. Alice and Bob generate raw key material by preparing and measuring states in the computational basis $\{\ket{n}\}$, while the F-qubit states are used exclusively for parameter estimation. By measuring error statistics in the F-qubit basis, Alice and Bob indirectly infer the phase error rate associated with the computational basis, thereby bounding Eve’s accessible information under collective (and, by extension, coherent) attacks. Importantly, this enables unconditional security despite the reduced dimensional support of the checking states.

A key result of this scheme is that it preserves the principal advantages of HD-QKD. The secret key rate per sifted photon scales as
\begin{equation}
R = \log_2 d - h_d(E_d) - h_d(E_d^{\mathrm{ph}}),
\label{freq}
\end{equation}
where $E_d$ and $E_d^{\mathrm{ph}}$ are the dit and phase error rates, respectively, and $h_d$ is the $d$-dimensional Shannon entropy. Consequently, the protocol achieves information densities exceeding one bit per detected photon while maintaining the dimension-dependent increase in tolerable error rates characteristic of high-dimensional systems.

Experimentally, Scarfe \emph{et al.} demonstrated this protocol in a noisy laboratory Free-space channel using orbital angular momentum (OAM) modes of light in $d=4$. By exploiting the larger spatial extent and improved mode overlap of F-qubit states relative to conventional Fourier modes, they achieved a measured sifted key rate of $R \approx 1.28$ bits per photon under realistic noise conditions. Fig.s~1–3 illustrate representative F-qubit mode structures, the experimental OAM implementation, and the associated probability outcome matrices used for phase-error estimation.

From a systems perspective, the F-qubit protocol significantly reduces state-preparation and detection complexity. Because each checking state involves only two modes regardless of $d$, the experimental overhead does not scale with dimensionality, making this approach particularly attractive for spatial-mode, time-bin, and integrated photonic implementations. This qubit-like HD-QKD architecture therefore provides a practical route toward high-rate, noise-tolerant quantum key distribution in bandwidth-limited Free-space and satellite channels.

\begin{figure*}[htbp]
    \centering
    \includegraphics[width=1\linewidth]{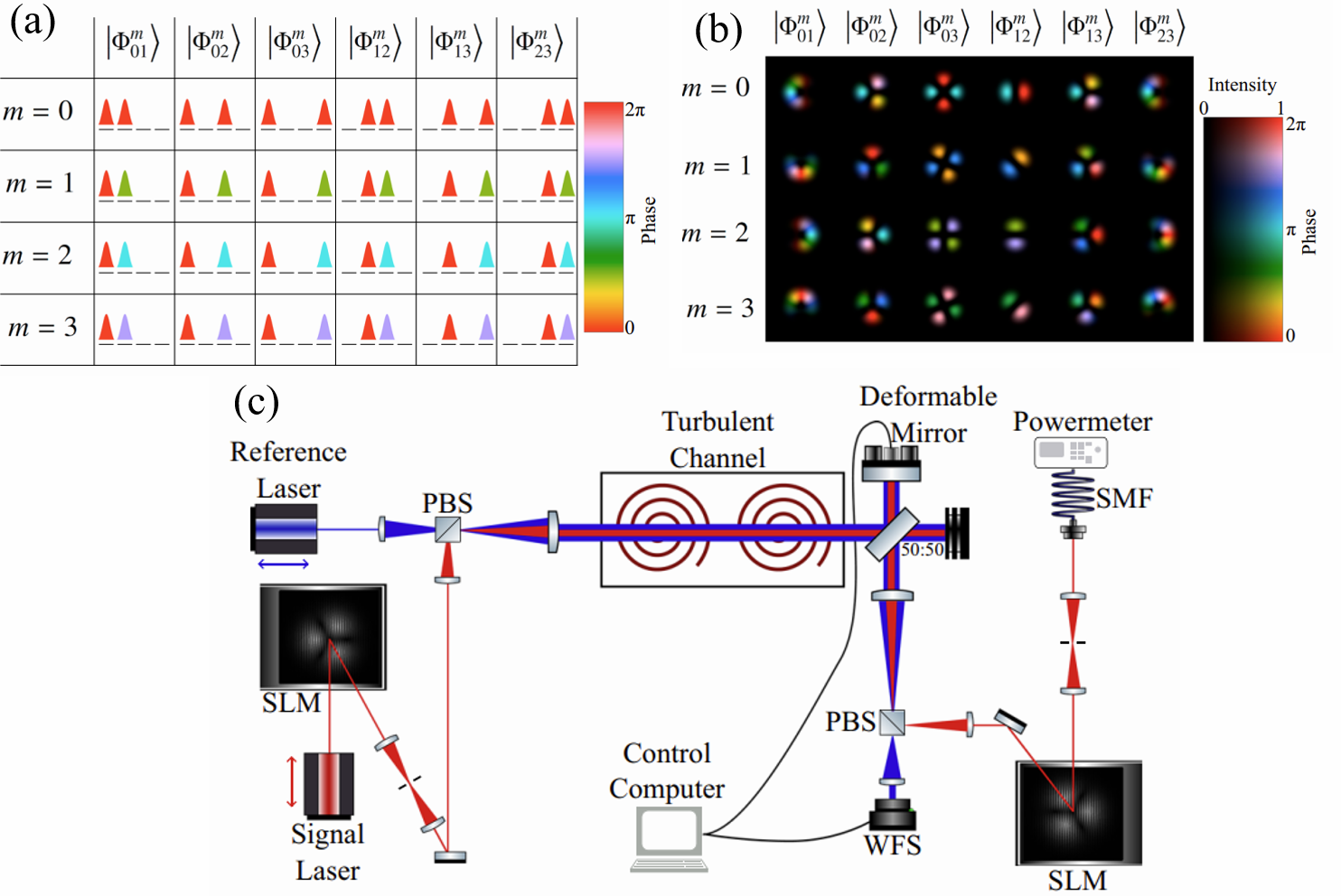}
    \caption{High-dimensional quantum key distribution (HD-QKD) using qubit-like states (Fourier-qubits). 
(a) Conceptual illustration of Fourier-qubit (F-qubit) states in a $d=4$ time-bin encoding. Logical basis states occupy distinct time bins, while F-qubits are constructed as equal-weight superpositions of only two logical states with a discrete relative phase $\omega_d^m = e^{2\pi i m/d}$, enabling phase-error estimation without full mutually unbiased bases. 
(b) Experimental realization of the F-qubit protocol using orbital angular momentum (OAM) modes of light in a noisy Free-space channel. Spatial light modulators (SLMs) generate and project qubit-like superpositions of Laguerre--Gaussian modes, while adaptive optics compensate turbulence-induced distortions prior to detection. 
(c) Measured probability outcome matrices for the F-qubit basis in four dimensions, showing the ideal theoretical distribution (left) and experimentally observed distribution after propagation through a turbulent channel (right). These statistics enable indirect reconstruction of the phase error rate and demonstrate secure key generation exceeding one bit per sifted photon\cite{scarfe2025fast}}
    \label{fig:hdqkd2}
\end{figure*}

\begin{figure}
\centering
\includegraphics[width=1\columnwidth]{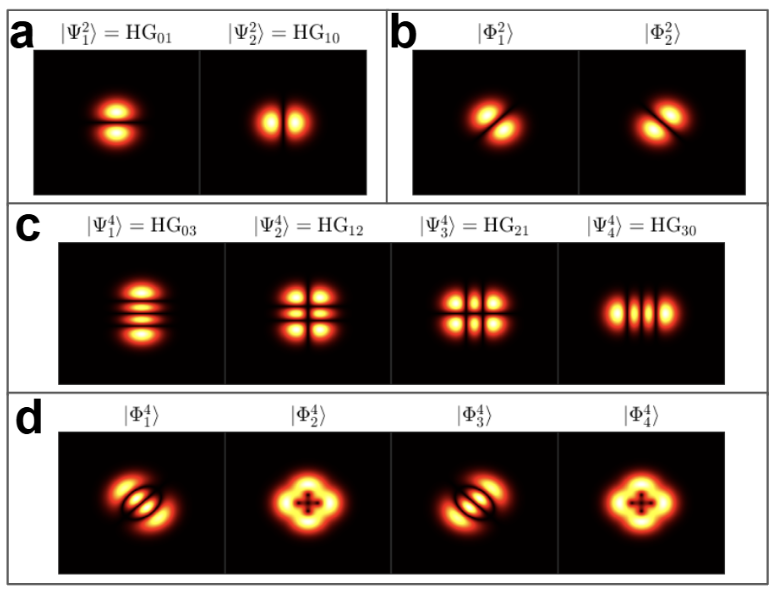}
\caption{Simulated amplitudes of Hermite Gauss bases of dimensions  \(d=2\) (a) and  \(d=4\) (c) as well as their MUBs (b,d) \cite{meyer2025analogy}.}
\label{fig:modes}
\end{figure}

In HD QKD, the sender (Alice) and receiver (Bob) use a set of $d$ orthogonal states (modes), with mutually unbiased bases (MUBs) $\{|\psi_i\rangle\}_{i=1}^d$, satisfying
\begin{align}
    \langle \psi_i | \psi_j \rangle &= \delta_{ij}, \\
    |\langle \phi_i | \psi_j \rangle|^2 &= \frac{1}{d}, \quad \forall i,j \in \{1, \ldots, d\}
\end{align}
where $\{|\psi_i\rangle\}$ and $\{|\phi_i\rangle\}$ are two MUBs \cite{mafu2013higher}.

The mutual information between Alice and Bob for a $d$-dimensional protocol with fidelity $F$ (probability of no error) is given by Eq.~(9) in~\cite{cerf2002security}:
\begin{equation}
    I_{AB} = \log_2 d + F \log_2 F + (1-F) \log_2 \left( \frac{1-F}{d-1} \right)
    \label{eq:IAB}
\end{equation}
where $F = 1 - Q$ and $Q$ is the quantum bit error rate (QBER).

This equation generalizes the standard binary mutual information to higher dimensions and explicitly incorporates both the dimension $d$ and the fidelity.

A particularly powerful implementation of HD-QKD uses spatial modes of light, especially those carrying orbital angular momentum (OAM) \cite{yao2011orbital}. Spatial modes form an orthogonal basis in the transverse spatial profile of photons, enabling encoding in a high-dimensional Hilbert space \cite{vagniluca2020efficient}.

Spatial modes, such as Hermite-Gauss (HG) or Laguerre-Gauss (LG), provide scalable orthogonal states \cite{saleh2019fundamentals}:
\begin{align}
    \mathrm{HG}_{mn}(x,y) &= H_m\left(\sqrt{2}\frac{x}{w(z)}\right) H_n\left(\sqrt{2}\frac{y}{w(z)}\right) \exp\left(-\frac{x^2 + y^2}{w^2(z)}\right)
\end{align}
where $H_m$ is the Hermite polynomial and $w(z)$ is the beam waist.

MUBs for spatial modes can be constructed using the Fourier relation:
\begin{equation}
    |\Phi_n^d\rangle = \frac{1}{\sqrt{d}} \sum_{k=0}^{d-1} \exp\left(\frac{2\pi i n k}{d}\right) |\Psi_k^d\rangle
\end{equation}

Phase-only spatial light modulators (SLMs) are commonly used for preparation and measurement of these states.

For perfect fidelity ($F = 1$), the mutual information between Alice and Bob reaches $I_{AB} = \log_2 d$ bits per sifted photon, demonstrating the fundamental advantage of HD-QKD in information capacity. Moreover, the maximal tolerable quantum bit error rate (QBER) increases with dimensionality: for $d = 2$ the threshold is $Q_c \approx 14.64\%$, for $d = 3$ it rises to $Q_c \approx 21.13\%$, and it continues to increase for higher dimensions~\cite{cerf2002security}. This enhanced error tolerance allows HD-QKD systems to maintain secure key generation even under noisier or more turbulent channel conditions \cite{kamran2024induced}.

However, these benefits come with practical trade-offs. As the dimension $d$ increases, modal cross-talk and alignment errors become more significant, while higher-order spatial modes exhibit greater beam divergence and sensitivity to optical aberrations. These effects can reduce the achievable key rate and limit the maximum transmission distance. Consequently, optimal HD-QKD implementations \cite{bouchard2018experimental} must balance the dimensionality of the encoding basis against system stability, channel fidelity, and available detection resources to maximize both security and performance.




\section{Advances in Error Correction Protocols}

Quantum key distribution (QKD) systems must reconcile errors arising from photon loss, detector imperfections, and environmental noise to produce an identical shared secret key between the legitimate users, Alice and Bob. Because classical error correction inevitably leaks partial information to an eavesdropper, modern QKD protocols integrate error reconciliation and privacy amplification as a unified \emph{post-processing} pipeline that determines the final secret-key length and security proof \cite{terhal2015quantum}.

\subsection{Classical Reconciliation Protocols}

The earliest reconciliation method, \emph{Cascade}, employs interactive parity checks across multiple passes to identify and correct mismatched bits. While its bit-error efficiency $\beta$ can approach $0.99$, the multi-round communication overhead limits throughput in long-distance and satellite-based links, where latency is non-negligible. 

To overcome this, one-way forward error-correction (FEC) schemes using \emph{Low-Density Parity-Check} (LDPC) or \emph{polar} codes have become the standard. LDPC codes permit near-real-time decoding using belief-propagation algorithms and achieve efficiencies $\beta\!\approx\!0.95$–$0.98$ for typical QBER values (1–5\%). FPGA and ASIC implementations of LDPC decoders now operate at tens of megabits per second, enabling real-time key distillation even during short satellite passes of a few hundred seconds. Polar codes, which exploit channel polarization, have likewise been shown to approach the Shannon limit with reduced memory requirements, making them attractive for embedded QKD payloads.

\subsection{Continuous-Variable Error Correction}

In continuous-variable (CV) QKD, Alice and Bob exchange correlated Gaussian variables rather than discrete bits. Here, reconciliation—also termed \emph{information reconciliation}—requires mapping continuous samples to discrete codewords. Multidimensional reconciliation combined with multi-edge LDPC codes provides high efficiency ($\beta>0.9$) even under low signal-to-noise conditions. Real-time implementations use iterative belief-propagation decoders that exploit soft information from the channel, allowing CV-QKD to operate over metropolitan and satellite links despite excess noise and fading.

\subsection{Privacy Amplification and Finite-Key Analysis}

After reconciliation, \emph{privacy amplification} removes any residual information that may have leaked to an eavesdropper. This is achieved by applying universal hash functions (e.g., Toeplitz or Reed–Solomon matrices) to compress the reconciled key. For finite data blocks, as encountered in satellite passes, the final key length $l$ satisfies
\[
l = s_{\mathrm{raw}}[1 - h(Q)] - \mathrm{leak}_{\mathrm{EC}} - \Delta_{\mathrm{sec}},
\]
where $s_{\mathrm{raw}}$ is the number of sifted bits, $Q$ is the measured QBER, $\mathrm{leak}_{\mathrm{EC}}$ quantifies disclosed information during error correction, and $\Delta_{\mathrm{sec}}$ accounts for finite-sample statistical deviations. Finite-key analysis ensures composable security even for short-duration quantum links, such as those from a single satellite pass lasting a few minutes.

\subsection{Emerging Trends}

Recent developments focus on adaptive and hardware-accelerated error-correction frameworks. Machine-learning-assisted decoders dynamically tune LDPC parameters in response to fluctuating channel conditions, reducing reconciliation failure rates in turbulent or fading channels. Hybrid FEC architectures combining GPU-based LDPC decoders with FPGA privacy-amplification engines have been demonstrated to exceed 100 Mb/s post-processing throughput in laboratory QKD systems. In spaceborne implementations, radiation-hardened FPGAs executing on-board LDPC decoding minimize classical downlink bandwidth requirements and support fully autonomous satellite QKD operations \cite{stathis2024toward}.

Altogether, these advances in error-correction coding, finite-key security, and adaptive post-processing constitute a crucial enabler for next-generation Free-space and satellite-based quantum communication, bridging the gap between laboratory demonstrations and globally scalable quantum networks.

\section{Recent Advances in Fiber-Based Quantum Key Distribution}

The development of quantum key distribution (QKD) has progressed rapidly with advancements in photon source and detector technologies. The first entanglement-based QKD experiment to surpass the 100~km barrier employed the BBM92 protocol using time-bin entangled photon pairs. This system integrated superconducting single-photon detectors (SSPDs) based on NbN nanowires, optimized for high-speed detection at telecom wavelengths (1.5~$\mu$m), alongside a periodically poled lithium niobate (PPLN) waveguide as a high-brightness entangled photon source. To ensure phase coherence, stable planar lightwave circuit Mach–Zehnder interferometers (PLC-MZIs) were used. This work marked a substantial improvement over earlier entanglement-based QKD systems, extending viable transmission distances from tens to over a hundred kilometers through superior source brightness and detection efficiency~\cite{Honjo:08}.

Shortly thereafter, continuous-variable QKD (CVQKD) was experimentally demonstrated over 24.2~km of optical fiber, reaching a secure key rate of 3.45~kbps. The system deployed polarization multiplexing and frequency translation techniques to transmit a continuous-wave local oscillator (CW-LO) while mitigating guided acoustic wave Brillouin scattering (GAWBS) by more than 27~dB. Although performance was limited by reconciliation efficiency under low signal-to-noise conditions, the result showcased CVQKD’s potential for high-speed secure communications over metropolitan-scale distances~\cite{Xuan:09}.

Subsequent progress led to the demonstration of QKD over 90~km in parallel with bidirectional 1.25~Gb/s classical data traffic. This was accomplished using decoy-state BB84 with sub-nanosecond gated InGaAs avalanche photodiodes (APDs) and temporal filtering to suppress Raman noise. The system achieved 507~kbps at 50~km and 7.6~kbps at 90~km, and successfully coexisted with 10~GbE over 65~km, providing a significant step toward QKD integration with real-world fiber infrastructures~\cite{PhysRevX.2.041010}.

\begin{figure*}[htbp]
    \centering
    \scalebox{0.45}{\includegraphics{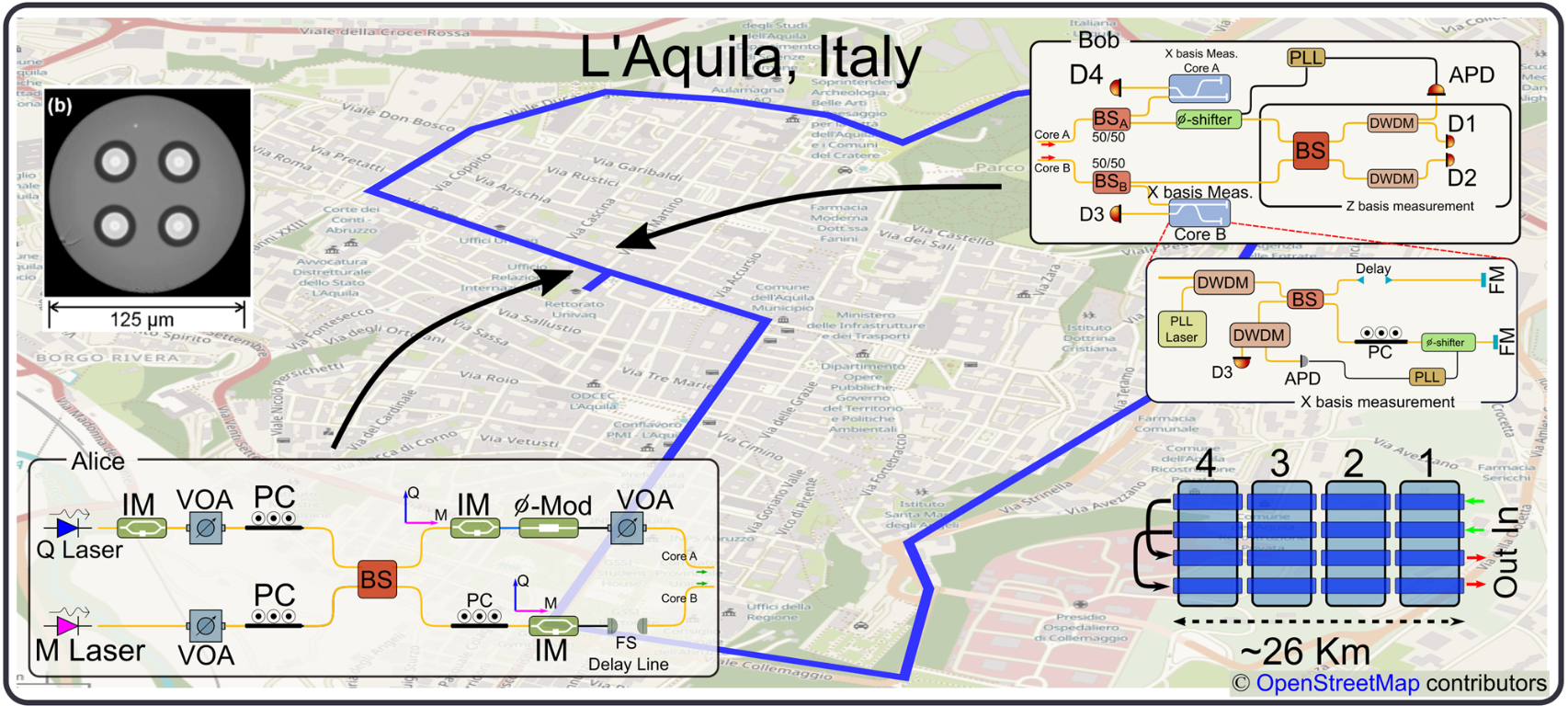}}
    \caption{Field-deployed high-dimensional QKD over multicore fiber. Hybrid time-path encoded states are transmitted through a 52~km loop formed by two spatial cores of a deployed four-core fiber. The system incorporates decoy-state modulation and active phase stabilization using a dual-band phase-locked loop. Adapted from: \textit{Nature Communications}, 2024.}
    \label{Fig_S3}
\end{figure*}

Further extending secure communication range, a record-setting Measurement Device Independent QKD (MDIQKD) implementation reached 404~km using ultralow loss fiber. This system employed an optimized four intensity decoy state protocol and high efficiency superconducting nanowire single photon detectors (SNSPDs), while accounting for finite size statistical effects. By removing trust in measurement devices, MDIQKD enhances security against detector based side channel attacks and confirms feasibility for future satellite compatible links~\cite{yin2016measurement}.
\begin{figure*}[htbp]
    \centering
    \scalebox{0.575}{\includegraphics{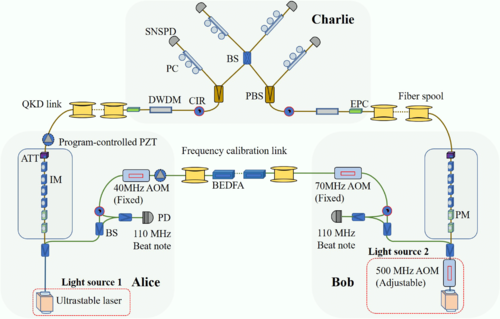}}
    \caption{Schematic of MDI-QKD over 404~km. Independent sources at Alice and Bob generate phase-randomized weak coherent states, which interfere at a central node (Charlie). Decoy-state techniques and advanced SNSPDs enable secure long-distance communication. Adapted from: Yin et al \cite{jiang2016pairing}}
    \label{Fig_S1}
\end{figure*}
Other efforts have focused on reducing system cost. One demonstration used light emitting diodes (LEDs) at 1310~nm instead of lasers, in combination with passive optical elements and a decoy state protocol, achieving 10.9~kbps secure key rate over 1~km. This low cost architecture illustrates the viability of QKD in local, metropolitan applications~\cite{Xia:19}.
The most advanced system to date combines high-coherence laser sources, heterodyne frequency calibration, and superconducting detectors to implement the Sending-or-Not-Sending Twin-Field QKD (SNS-TFQKD) protocol. Operating over 658~km of ultralow-loss fiber, this system achieved a secure key rate of $9.22 \times 10^{-10}$ bits per pulse (0.092~bps), confirmed by finite-size analysis. Notably, the system simultaneously served as a distributed vibration sensor, locating environmental perturbations with 1~km precision over a 500~km calibration link. These results push the frontier of QKD distance and illustrate its multifunctional potential in hybrid communication-sensing networks~\cite{PhysRevLett.128.180502}.

\section{Advances in Quantum Communication: Key Experimental Milestones }

\subsection{Practical Free-space Quantum Key Distribution over 1 km}

Buttler et al. (1998)  demonstrated a practical Free-space QKD system operating over a 1 km atmospheric path at night using the B92 protocol with non-orthogonal polarization states. The setup employed a 772 nm diode laser attenuated to ~0.1 photons per pulse, with polarization encoded via a Pockels cell and analyzed using single-photon detectors and spatial filtering to suppress background noise. Reliable bit generation was achieved with bit error rates below 1.5$\%$, validating secure key exchange under realistic outdoor conditions. The study also assessed the feasibility of ground-to-satellite QKD, estimating achievable nighttime key rates of 35–450 Hz with standard optics, and projecting significant improvements using adaptive optics. These results underscore the practical potential of QKD for long-distance, line-of-sight applications, including satellite rekeying. Their system achieved successful QKD over Free-space paths up to 950 m with low bit error rates. At 950 m and 0.1 photons per pulse, Bob detected ~50 bits/s, matching theoretical predictions. QBER was 0.7$\%$ percent at 240 m and 1.5$\%$ at both 500 m and 950 m. A two-dimensional parity check allowed generation of error-free keys from the raw sequences. The system’s security was analyzed under two eavesdropping models intercept resend and beam splitter attacks—and found to be robust, with privacy amplification able to counteract minimal information leakage. These results support the feasibility of secure Free-space QKD and suggest that, with enhancements like adaptive optics, such systems could be extended to satellite communications with realistic key rates.

\subsection{Practical Free-space quantum key distribution over 10 km in daylight and at night}

Hughes \textit{et~al.} (2002) demonstrated the feasibility of free‐space QKD over a \(10\;\mathrm{km}\) atmospheric link under both daylight and nighttime conditions using the BB84 protocol.  The system achieved cryptographic‐quality key generation despite background noise, and a generalized performance model was developed to extrapolate behaviour under varying atmospheric and instrumental conditions—suggesting QKD could be extended to \(45\;\mathrm{km}\) at night.  Their results provide a foundation for scalable, line-of-sight quantum-secure communication systems.

\begin{figure*}[htbp]
    \centering
    \scalebox{0.4}{\includegraphics{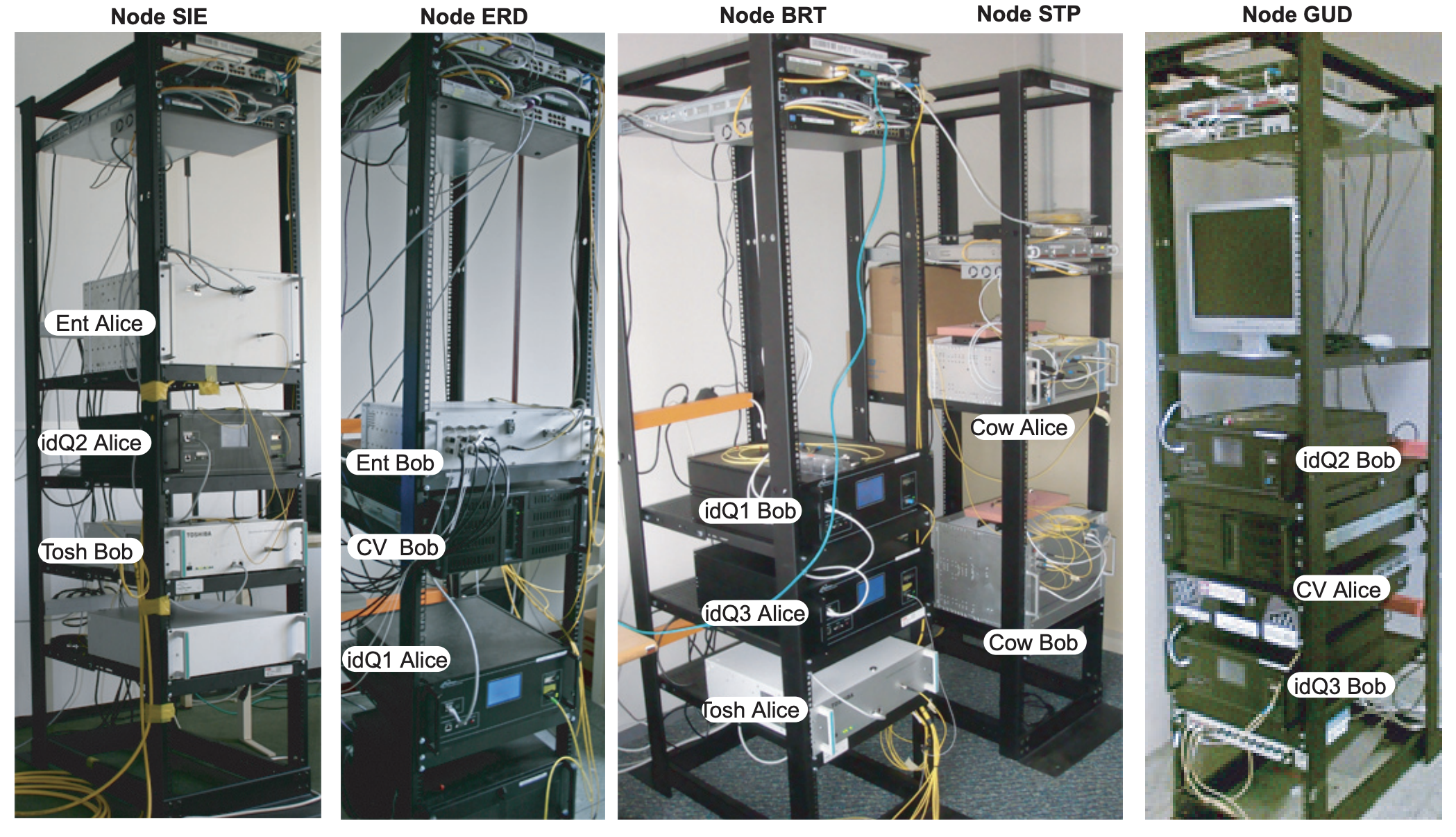}}
     \caption{Photographs of the SECOQC quantum key distribution (QKD) network node racks deployed in the Vienna metropolitan field trial.  Each 19\,in rack integrates polarization- and phase-encoded QKD modules from multiple vendors, key-management units, classical authentication hardware, optical patch panels, and redundant power supplies, enabling plug-and-play interconnection of heterogeneous quantum links within the city-wide SECOQC testbed.  Reproduced from \cite{peev2009secoqc}.}
    \label{Fig_SECOQC}
\end{figure*}

\medskip
\noindent
The transmitter (Alice) employed four \(772\;\mathrm{nm}\) diode lasers, each encoding one of the BB84 polarization states.  A \(1\;\mathrm{MHz}\) clock triggered a \(1550\;\mathrm{nm}\) timing pulse followed by a \(1\;\mathrm{ns}\) data pulse, attenuated to \(<1\) photon on average.  Spectral and spatial filtering suppressed background light before transmission.

\medskip
\noindent
The receiver (Bob) used an \(18\;\mathrm{cm}\) Cassegrain telescope, a \(0.1\;\mathrm{nm}\) interference filter, and passive polarization analysis with single-photon detectors (SPDs).  As illustrated schematically in Fig.~\ref{Fig_fig_turbo1}, photons were routed through beam splitters to polarization analysers corresponding to the rectilinear and diagonal bases; detections were registered within narrow \(\sim1\;\mathrm{ns}\) timing windows synchronized to the transmitted pulse.

\medskip
\noindent
The link spanned \(9.81\;\mathrm{km}\) between elevated sites in New Mexico, with visual-alignment cameras and a wireless public-channel link supporting operation.  The system performed reliably across changing conditions:

\begin{itemize}
  \item \textbf{Daylight} (\(\langle\mu\rangle \approx 0.49\)): sifted key rates of \(100\)–\(2000\;\mathrm{bits/s}\) with a bit-error rate (BER) of \(\sim5\%\).
  \item \textbf{Night} (\(\langle\mu\rangle \approx 0.14\)): comparable sifted rates with a reduced BER of \(\sim2\%\).
\end{itemize}

\noindent
Background light dominated the noise budget during the day, whereas detector dark counts were limiting at night.  After key sifting, error correction, and privacy amplification, final secret keys were produced with a secrecy efficiency up to \(8\times10^{-4}\) secret bits per transmitted bit.  In total, more than \(1.68\times10^{5}\) secret bits were extracted over the combined day–night trials, and all keys passed standard cryptographic randomness tests.

\medskip
\noindent
The system was shown to be secure against realistic intercept–resend and photon-number-splitting attacks, underscoring the practical viability of Free-space QKD in real-world environments \cite{clivati2022coherent}.

\subsection{Work of the DARPA quantum network}

The DARPA Quantum Network, developed in 2005 through a collaboration between BBN Technologies, Harvard University, and Boston University, was the first operational QKD network. Unlike earlier standalone QKD systems, this network provided continuous, real-world key distribution across metropolitan-scale distances using both fiber and Free-space optical links. Its goal was to demonstrate the feasibility of quantum-secure communication on a networked scale, capable of supporting multiple QKD technologies and operating reliably under practical conditions. The network spanned up to 29 km across Cambridge, Massachusetts, using standard SMF-28 fiber and incorporated six nodes, including Alice, Bob, Anna, and Boris, with programmable optical switching between them. It integrated several QKD platforms: BBN’s Mark 2 weak-coherent phase-modulated fiber system, a polarization-entangled photon system developed with Boston University, and high-speed Free-space systems contributed by NIST and QinetiQ. The QKD protocols implemented included BB84, error correction via Cascade and Niagara, entropy estimation, privacy amplification, and IPsec-based authentication. The software stack supported a range of entropy models—including Bennett, Slutsky, Myers-Pearson, and Shor-Preskill—as well as different sifting modes, such as classic BB84 and SARG. Performance varied by link configuration. For example, pulses with a mean photon number of 0.5 transmitted from Anna to Bob yielded approximately 1,000 secret bits per second with a quantum bit error rate (QBER) of ~3$\%$. In contrast, the Boris link suffered from high attenuation (11.5 dB) and inefficient detectors, requiring temporary operation at µ = 1.0 and ultimately resulting in zero secret key yield. The Mark 2 system demonstrated stable operation at a rate of 3.3 million pulses per second, with adjustable photon rates and attenuation settings to match fiber span losses. The system’s design is shown in Fig. \ref{Fig_SECOQC}. Fiber lengths and attenuation posed practical challenges such as connector-induced losses and spans with effective path lengths exceeding 50 km, which required careful calibration of photon rates and attenuation settings. The network topology incorporated both fiber and Free-space links, and included partially operational entangled photon nodes (Alex and Barb), with plans for further expansion using additional QinetiQ hardware. Operational as of early 2005, the DARPA Quantum Network demonstrated the practical viability of QKD in a multi-node, metropolitan-scale setting, laying a critical groundwork for future scalable quantum-secure communications \cite{pan2024evolution}. Software enhancements, such as the Niagara forward error correction (FEC) protocol, significantly reduced communication overhead and CPU usage compared to Cascade, albeit with a modest tradeoff in coding efficiency. Ongoing efforts focused on activating entangled and Free-space links, deploying improved hardware like superconducting detectors, and refining communication protocols. Ultimately, the project illustrated how robust cryptographic services could be sustained over complex real-world networks, even in the face of hardware variability, attenuation, and environmental challenges \cite{javadpour2023encryption}.

\subsection{The SECOQC quantum key distribution network in Vienna}

As part of the SECOQC project, a Free-space QKD system was developed by the University of Munich in 2009 to evaluate the feasibility of secure, last-mile quantum communication over a line-of-sight urban link. Designed for integration with fiber-based nodes in metropolitan QKD networks, the system demonstrated robust, high-rate key exchange using polarization-encoded weak coherent pulses (WCPs). It was based on the BB84 protocol with decoy states and employed attenuated 850 nm laser pulses to encode information in four polarization states. The sender (Alice), located at the Siemens Forum (node FRM), used laser diodes to generate weak pulses with varying polarization and photon number. The receiver (Bob), located in a neighboring building (node ERD), detected incoming photons using silicon avalanche photodiodes (Si-APDs). A telescope collected the Free-space beam, which then passed through a series of optical and spectral filters designed to suppress background noise. Narrow-bandpass filters, spatial filtering, and strict alignment maintenance via a real-time feedback enabled reliable operation in various lighting conditions. Key generation rates exceeded 10 kbit/s, surpassing typical short-distance fiber-based QKD systems, and remained consistent across day–night cycles.

Integration with SECOQC node modules allowed the Free-space link to connect seamlessly with the broader fiber-based QKD infrastructure. By contributing secure key material to the SECOQC trusted repeater layout, the system demonstrated that Free-space links could serve as effective last-mile access channels, particularly in settings where fiber deployment was impractical or cost-prohibitive. The setup supported fully autonomous, continuous 24/7 operation, maintaining low quantum bit error rates (QBER) and demonstrating strong resilience to environmental fluctuations. The SECOQC Free-space QKD implementation confirmed that polarization-based BB84 QKD could be performed reliably over short, line-of-sight urban links in real-world conditions. Its high key rates, stable performance, and smooth integration into the existing network highlighted its practicality as a flexible access solution for metropolitan-scale quantum-secure communications and emphasized the interoperability of heterogeneous QKD technologies.

\begin{figure*}[htbp]
    \centering
    \scalebox{0.6}{\includegraphics{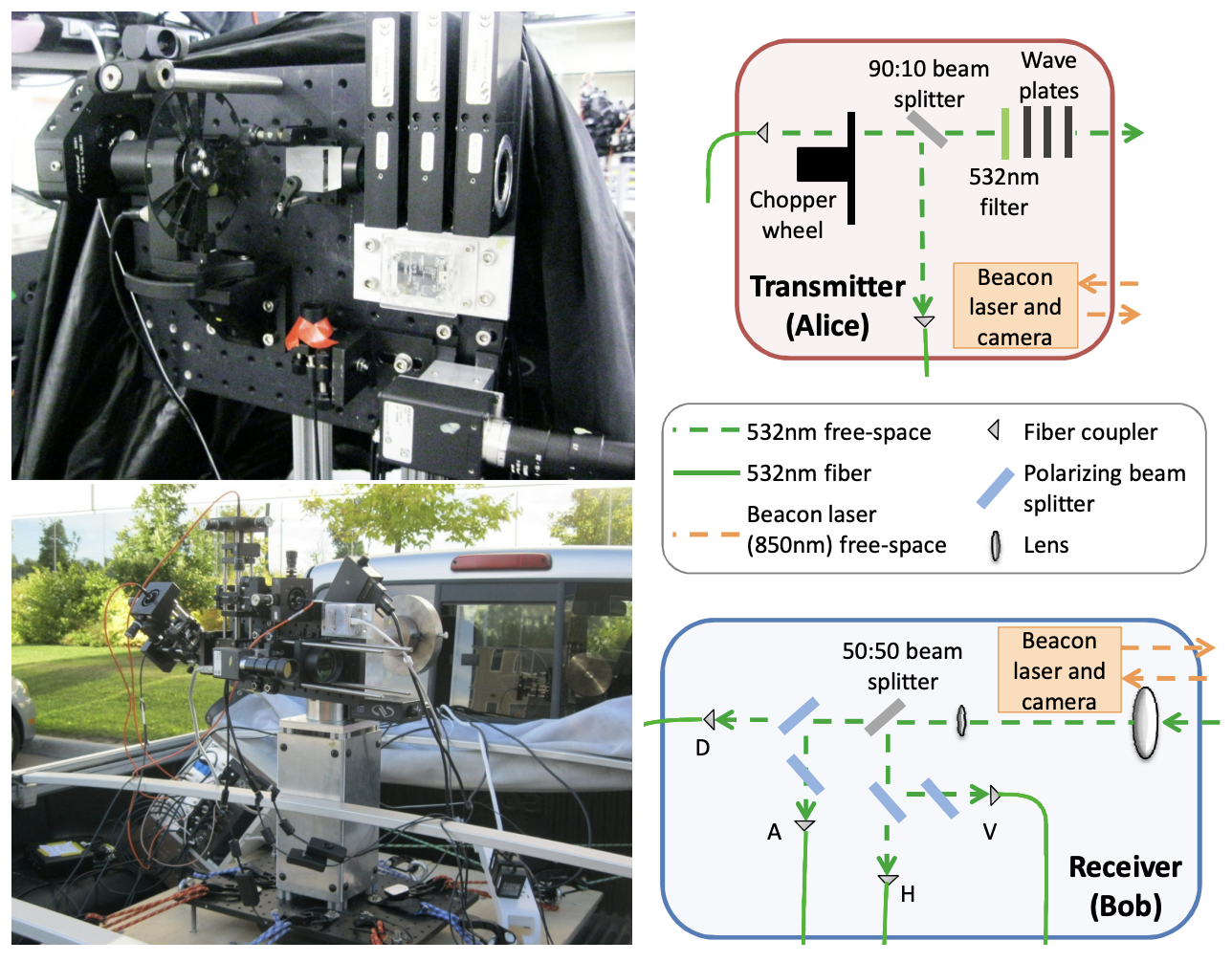}}
     \caption{Transmitter (Alice, top) and receiver (Bob, bottom) setups for Free-space QKD with a moving receiver. Alice produces a $\sim$10\,mm collimated beam using a chopper wheel embedded with polarization films to encode the BB84 states. A 10\% reflective beam splitter diverts a portion of the light to a fiber-coupled polarimeter for real-time polarization state tomography. The measured data drives a set of motorized wave plates to compensate polarization drifts caused by the transmitter fiber. Only pulses passing through open chopper slots are used for key generation. Bob, mounted on a moving platform, implements a passive basis-choice measurement using a beam splitter and two polarization analyzers. Adapted from Bourgoin \textit{et al}\cite{bourgoin2015free}.}
    \label{Fig_Optics_Express}
\end{figure*}

\subsection{Free-space quantum key distribution to a moving receiver}

Bourgoin et al. (2015) reported the first successful demonstration of Free-space QKD from a stationary transmitter to a moving receiver simulating the angular velocity of a low-Earth-orbit (LEO) satellite. The experiment addressed key engineering challenges for satellite-based QKD, including real-time beam tracking, polarization drift compensation, and time-of-flight correction. They validated the feasibility of secure key exchange under dynamic conditions representative of satellite passes. The sender (Alice) was located in a laboratory and generated 532 nm weak coherent pulses using sum-frequency generation from an 810 nm pulsed laser and a 1550 nm continuous-wave laser. These pulses were modulated in polarization and intensity to implement the BB84 protocol with decoy states. The beam was transmitted through a telescope and stabilized using a beacon laser and a camera-based tracking system. To correct for real-time polarization drift caused by fiber-induced birefringence, the transmitter incorporated a polarization compensation module with motorized waveplates guided by a tomographic chopper system. The receiver (Bob) was mounted on a pickup truck traveling at 33 km/h, matching the apparent angular speed of a satellite in a 600 km LEO orbit. Bob passively analyzed photon polarization using beam splitters, waveplates, and single-photon detectors. Acquisition and tracking were maintained via 850 nm beacons and a real-time feedback system. Fig. 1 illustrates the experimental layout and truck path, and Fig. 2 details the optical components used by Alice and Bob. Despite the motion of the receiver, the quantum link remained stable at an angular rate of 0.75°/s, exceeding typical LEO satellite speeds. 

The system experienced a total link loss of 30.6 dB, attributed primarily to beam divergence and pointing jitter, particularly at the mobile receiver. After acquisition, angular deviation was reduced to 0.005° at Alice and 0.06° at Bob. Time-of-flight (TOF) variations were corrected using GPS-based models, and QBER and photon count rates were monitored throughout the transmission. During a 4-second high-SNR window, the system achieved a raw key rate of approximately 11.5 kb and a secure key length of 160 bits under asymptotic assumptions. The dominant source of error was a QBER of ~6$\%$, mainly due to imperfections in source fidelity and polarization state purity. These were actively compensated in real time, with remaining errors traced to beam asymmetry and modulator limitations, as identified through modeling. 

Their work demonstrated the practical viability of QKD between a ground station and a moving platform, offering critical insights for future satellite-based implementations. By integrating real-time polarization control, precision tracking, and accurate timing correction, the experiment successfully simulated a realistic satellite pass and generated secure keys. Although the secure bit yield was modest, the results confirmed that with further refinements, such as lowering intrinsic QBER and improving receiver stability, practical satellite QKD is within reach. The system layout and compensation strategies presented here offer a robust foundation for future ground-to-space quantum communication missions.

\subsection{Long-distance Free-space quantum key distribution in daylight towards inter-satellite communication}

Liao et al. (2017) conducted a 53 km Free-space QKD experiment using 1,550 nm photons, demonstrating the feasibility of secure Free-space quantum communication under high-loss and high-background conditions. The high feasibility was achieved by using telecom-band photons, upconversion detection, and single-mode fiber spatial filtering. Motivated by the need for inter-satellite quantum communication in daylight, the study laid critical groundwork for satellite-constellation-based global quantum networks.

The experiment was conducted across Qinghai Lake in China, with the sender (Alice) and receiver (Bob) separated by a distance of 53~km. Alice employed four distributed-feedback (DFB) lasers operating at a wavelength of 1,550.14~nm (see Fig.~\ref{fig:10}) and implemented the BB84 protocol with decoy states.

Signal photons were collimated using a 254~mm aperture telescope, achieving a near-diffraction-limited divergence angle of approximately 12~$\mu$rad. At Bob’s site, a 420~mm parabolic mirror collected the incoming photons, which were subsequently coupled into a single-mode fiber (SMF). Upconversion single-photon detectors (SPDs), operating at room temperature, converted the incoming telecom-band photons into visible wavelengths, enabling detection with silicon avalanche photodiodes (Si-APDs). A GPS-based timing system, together with a high-frequency optical tracking system, ensured synchronization and beam stability throughout the transmission.

Despite daylight conditions, the system operated stably with a total channel loss of 48 dB, comprising 14 dB from fiber coupling and 34 dB from geometric spreading, atmospheric absorption, and detector inefficiencies. Using low-density parity-check (LDPC) codes for error correction, the experiment achieved secure key rates ranging from 20-400 bits per second. Over an effective transmission period of 1,756 seconds, a total of 157,179 secure bits were generated. The average quantum bit error rate (QBER) remained below 3.5$\%$. The upconversion SPDs contributed significantly to performance, offering low dark count rates (~20 Hz) and strong filtering of background light. The 53 km link, operated under conditions comparable to or more challenging than those expected in low-Earth orbit (LEO) inter-satellite scenarios, validated the viability of daylight operation for satellite QKD. Their results also highlighted the compatibility between Free-space and fiber QKD systems, offering a scalable layout for future global quantum communication networks. Continued advancements in detector performance and beam pointing precision are expected to further improve system reliability and expand operational range.

\begin{figure*}[htbp]
    \centering
    \scalebox{0.3}{\includegraphics{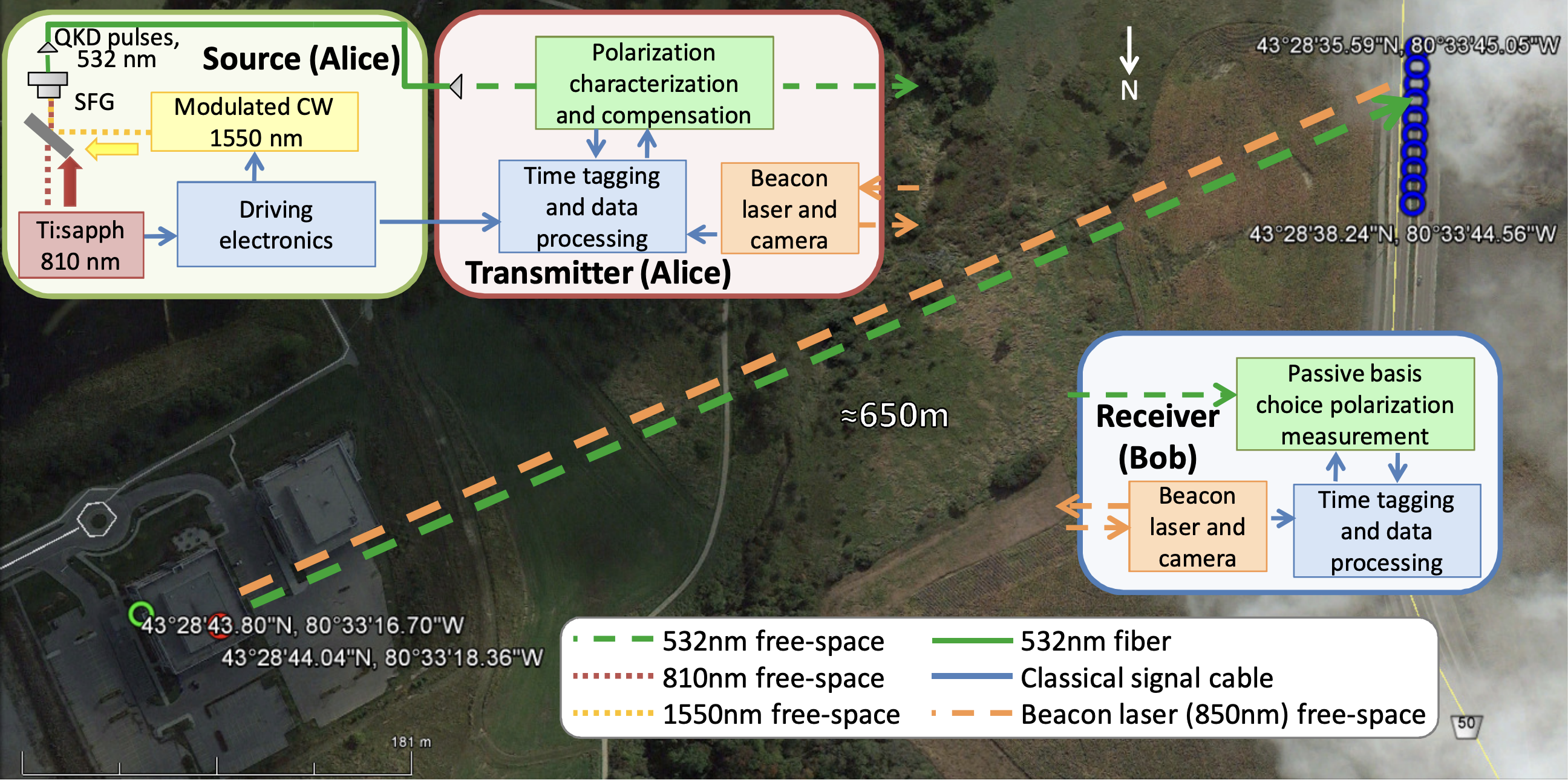}}
     \caption{Schematic overview of the moving-receiver Free-space QKD field test and corresponding map. Alice comprises a source (green circle) located in a laboratory on the building’s ground floor and a rooftop transmitter (red circle) fed via optical fiber. Bob, mounted on a truck, follows an $80\,\mathrm{m}$ road segment; blue circles mark the truck’s position at $1\,\mathrm{Hz}$ during the run. An active laser-beacon tracking and pointing system maintains link alignment while the truck is in motion, and a wireless LAN provides the authenticated classical channel. The line-of-sight separation between the transmitter and the truck is $\approx 650\,\mathrm{m}$. Map data: Google, DigitalGlobe. Reproduced from Bourgoin \textit{et al}\cite{bourgoin2015free}.}
    \label{Fig_Optics_Express2}
\end{figure*}

\subsection{Satellite-to-ground quantum key distribution}

Building on their previous work, Liao et al. (2017) demonstrated the first successful satellite-to-ground QKD, using the Micius satellite to achieve secure key exchange over distances of up to 1,200 km. This landmark experiment laid the foundation for a global-scale quantum communication network by overcoming the fundamental limitations of terrestrial QKD systems.

The \textit{Micius} satellite, operating in a \SI{500}{\kilo\meter} Sun-synchronous orbit, carried an \SI{850}{\nano\meter} decoy-state BB84 QKD transmitter comprising eight laser diodes encoding polarization states. A \SI{300}{\milli\meter} aperture telescope on board achieved a narrow beam divergence of approximately $\SI{10}{\micro\radian}$. 

On the ground, receiving operations were conducted at the Xinglong station using a \SI{1}{\meter} telescope equipped with BB84 decoding optics and four single-photon detectors. Maintaining precise alignment during orbital passes at a velocity of $\sim$\SI{7.6}{\kilo\meter/\second} required a sophisticated acquiring, pointing, and tracking (APT) system installed on both satellite and ground terminals. 

Synchronization between the satellite and ground station was facilitated by a \SI{532}{\nano\meter} pulsed beacon laser, achieving a timing jitter of $\sim$\SI{0.5}{\nano\second}. To preserve polarization fidelity throughout the satellite pass, a motorized half-wave plate enabled dynamic polarization compensation in response to orbital motion.

On December 19th, 2016, Micius successfully established a QKD link with the ground station at distances ranging from 645 km to 1,200 km. Over 273 seconds of transmission, the system registered 1.67 million sifted key bits, with key rates peaking at 12 kbit/s near the closest approach and decreasing to ~1 kbit/s at the farthest range. The average quantum bit error rate (QBER) was 1.1$\%$, consistent with expected contributions from background noise and polarization visibility. Fig. 3 presents key performance metrics from a single satellite pass, including orbital distance, real-time sifted key rate, and QBER, which varied with angular velocity, especially during overhead passes when tracking became more challenging. Following error correction and privacy amplification, the session yielded 300,939 final secure bits. Across 23 nights of testing, the system consistently achieved QBERs between 1–3$\%$ and peak secure key rates of up to 40.2 kbit/s. These results outperformed fiber-based QKD systems at comparable distances by more than 20 orders of magnitude in link efficiency, clearly demonstrating the unique advantages of satellite-based quantum communication.
By bypassing the severe photon losses inherent in fiber and terrestrial channels, the satellite-based approach enabled secure quantum key exchange over intercontinental scales. With Micius acting as a space-based relay, the experiment provided a scalable layout for interlinking metropolitan QKD networks.

\subsection{An integrated space-to-ground quantum communication network over 4,600 kilometers}

Chen et al. (2021) reported the first successful realization of a large-scale quantum communication network integrating satellite-based and terrestrial QKD systems. Spanning 4,600 km, the network combined over 700 fiber-based QKD links with two satellite-to-ground channels, enabling practical, secure quantum key distribution across multiple cities and remote regions. This achievement marked a critical step toward the development of a global quantum internet. The architecture comprised four metropolitan fiber QKD networks (QMANs), a 2,000 km national-scale fiber backbone, and two high-speed satellite–ground links. The terrestrial fiber QKD links employed the BB84 protocol with decoy states, using commercial InGaAs/InP and up-conversion single-photon detectors operating at 40 MHz and 625 MHz, respectively.

\begin{figure*}[htbp]
    \centering
    \scalebox{0.45}{\includegraphics{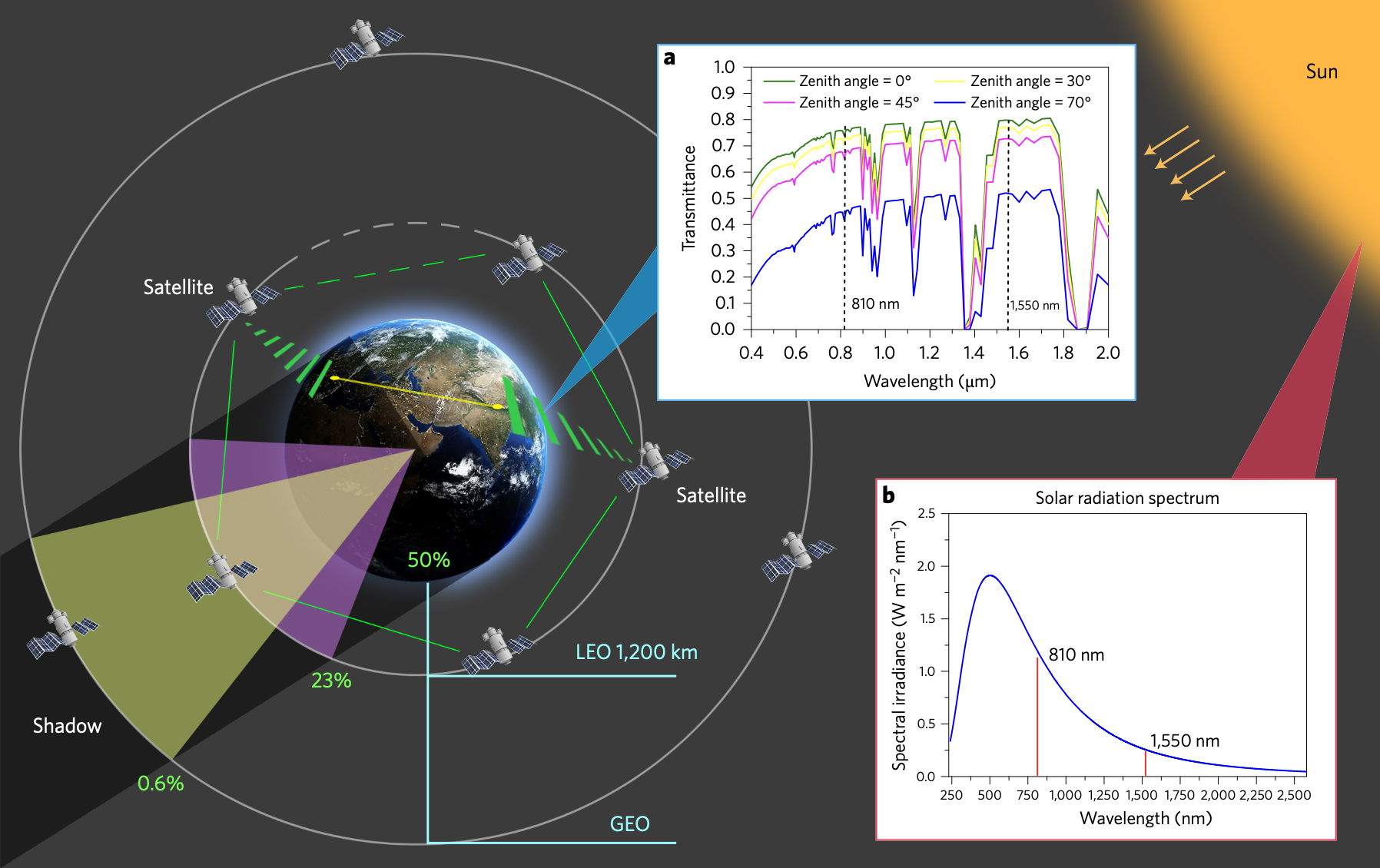}}
     \caption{Satellite-constellation-based global quantum network. A global quantum network requires many low-Earth-orbit (LEO) satellites or several geosynchronous (GEO) satellites to form a satellite constellation. The time a satellite spends in Earth’s shadow (“night”) is inversely proportional to its orbital height. (a) Atmospheric transmittance from the visible to the near-infrared at selected zenith angles. (b) Solar spectral irradiance from the visible to the near-infrared. Adapted from S.-K.~Liao \emph{et~al.} \cite{liao2017long}.}
    \label{Fig_NatPho1}
\end{figure*}

For satellite-based distribution, the Micius satellite transmitted BB84-encoded 850 nm photons at a 200 MHz repetition rate. Upgrades to ground station optics and spectral filters enhanced photon coupling and suppressed background noise, improving link performance. The satellite–ground channels connected the Xinglong and Nanshan stations, located 2,600 km apart, enabling end-to-end secure key exchange across the hybrid network. During a 364-second satellite pass, the system achieved a sifted key rate of up to 462 kbps and an average final secure key rate of 47.8 kbps—roughly 40 times higher than previous satellite QKD demonstrations.

\begin{figure*}[htbp]
    \centering
    \scalebox{0.5}{\includegraphics{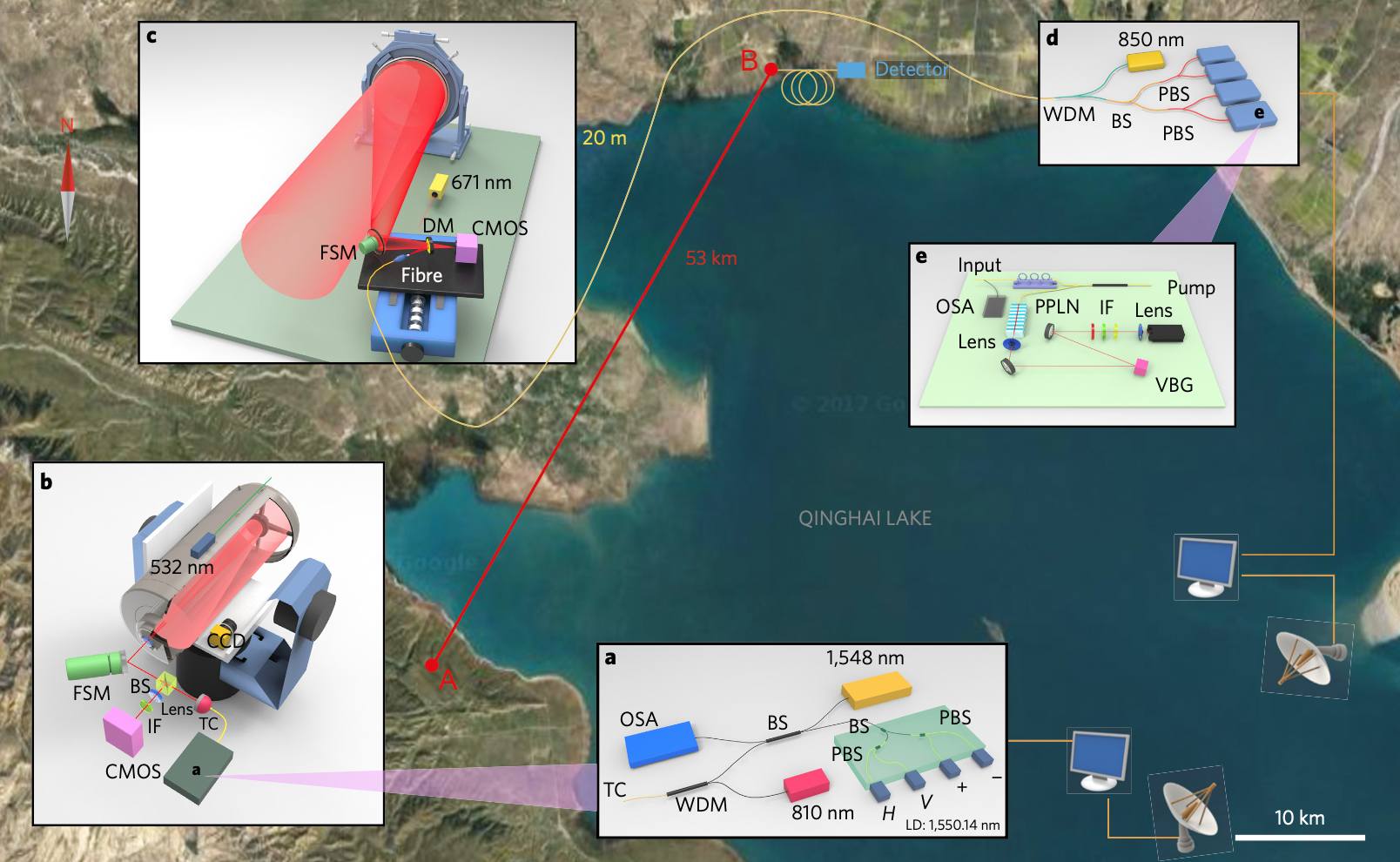}}
     \caption{Bird’s-eye view of the $53\,\mathrm{km}$ daylight Free-space QKD experiment around Qinghai Lake. Alice and Bob are positioned on opposite shores. (a) $1{,}550\,\mathrm{nm}$ laser diodes (LDs) are encoded into the four BB84 polarization states ($\lvert H\rangle$, $\lvert V\rangle$, $\lvert +\rangle$, $\lvert -\rangle$) using two polarizing beam splitters (PBSs) and one beam splitter (BS). An $810\,\mathrm{nm}$ beacon laser and a $1{,}548\,\mathrm{nm}$ reference laser are combined (WDM) and launched via a triplet collimator (TC) for optical alignment and tracking; an optical spectrum analyzer (OSA) calibrates the signal spectrum. (b) The sending terminal comprises a telescope on a two-axis rotation stage and an optical tracking system (CCD, fast-steering mirror, FSM; interference filter, IF). (c) The receiving terminal employs an off-axis parabolic mirror with single-mode-fiber (SMF) coupling. (d) Received photons are guided over a $20\,\mathrm{m}$ fiber to the measurement module with two PBSs, one BS, and four detectors. (e) Upconversion single-photon detector modules based on periodically poled lithium niobate (PPLN) use a narrow-band volume Bragg grating (VBG) to suppress background. For alignment and tracking details, see Methods. Map data: Google, CNES/Airbus, DigitalGlobe, Landsat/Copernicus. Adapted from S.-K.~Liao \emph{et~al.}\cite{liao2017long}.}
    \label{fig:10}
\end{figure*}

\begin{figure*}[htbp]
    \centering
    \scalebox{0.5}{\includegraphics{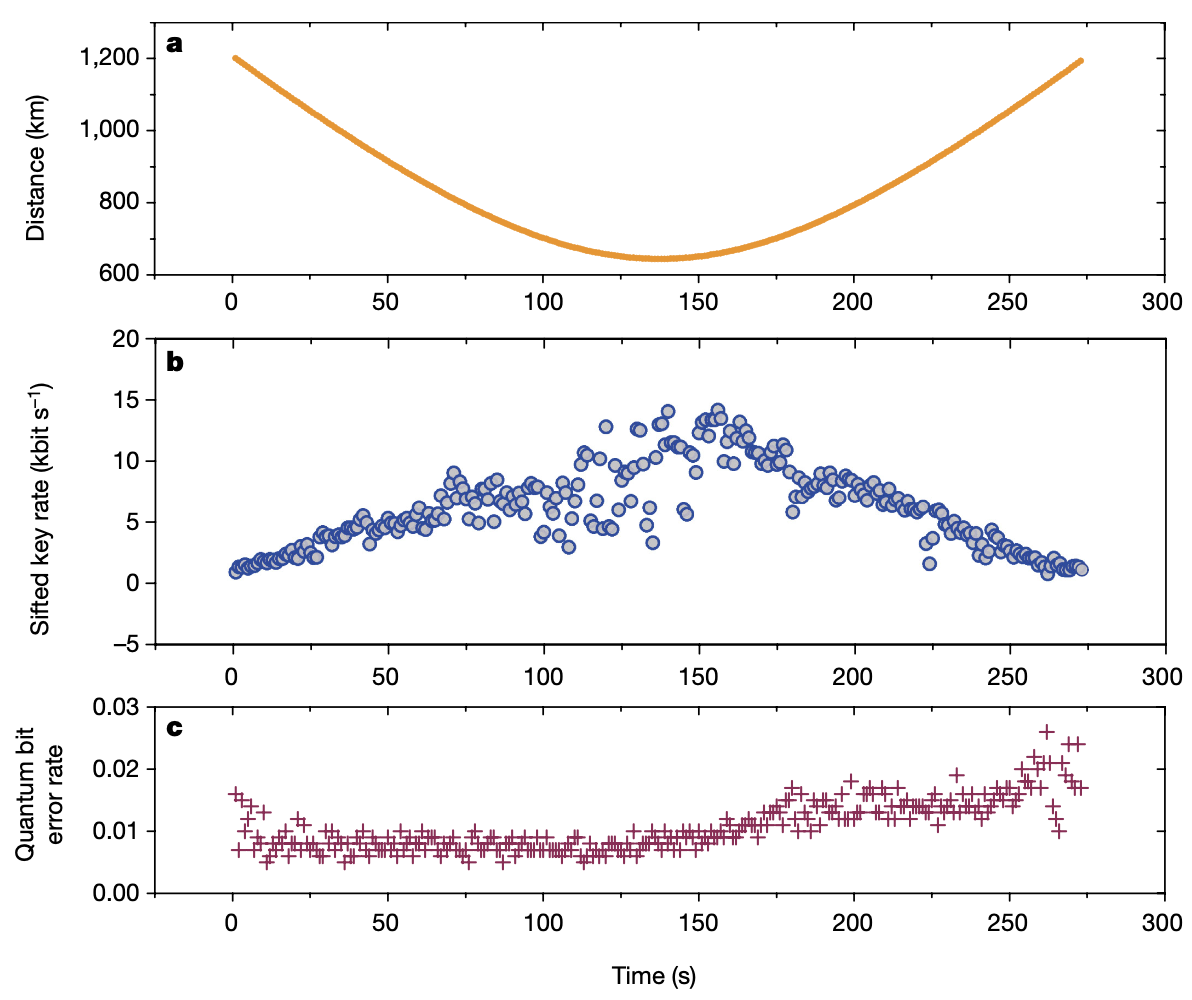}}
     \caption{Performance of satellite-to-ground QKD during a single orbit. (a) Trajectory of the \emph{Micius} satellite as measured from the Xinglong ground station. (b) Sifted key rate as a function of time and slant range (satellite–station distance). (c) Observed quantum-bit error rate (QBER). See main text for discussion; additional days are provided in Extended Data Table~2 and Extended Data Fig.~1. Adapted from S.-K.~Liao \emph{et~al}\cite{liao2017satellite}}
    \label{Fig_Nat1}
\end{figure*}

A total of 58.1 Mbit of sifted key material was collected, with an average QBER of just 0.5$\%$. High-performance tracking and adaptive exposure control enabled reliable operation at low elevation angles (~5°), extending the viable link distance beyond 2,000 km. Fig. 3 shows the sifted key rate and QBER versus distance (top), an illustration of the expanded satellite coverage angle (middle), and successful key generation up to 2,043 km (bottom).

Their results aligned with expected optical losses for geosynchronous satellite links, confirming the feasibility of long-distance satellite QKD. By successfully linking terrestrial fiber infrastructure with high-throughput satellite channels, the network demonstrated secure, high-rate QKD across continental scales. Addressing key challenges in scalability, hardware performance, and long-range operation, the experiment advanced the practical realization of a global quantum internet. Its demonstrated capabilities also support the feasibility of future satellite constellations and intercontinental QKD via geosynchronous orbits

\subsection{Free-space quantum key distribution during daylight and at night }

Cai et al. (2024) presented a comprehensive experimental demonstration of Free-space QKD over a 20 km terrestrial link operating continuously during both daylight and nighttime. By integrating a high-speed, stable light source with near-theoretical-limit noise suppression techniques, they addressed key challenges that had previously restricted QKD to nighttime and delayed post-processing. This work established a critical foundation for enabling continuous, all-day satellite-based quantum communication.

The experiment was carried out between a satellite payload prototype (Alice), stationed at the Silk Road Resort (altitude 2266~m), and the Nanshan ground station (altitude 2070~m). The system implemented a 625~MHz BB84 decoy-state protocol using a Sagnac-interferometer-based modulation scheme for polarization and intensity control. Quantum photons at 1550~nm were filtered through a Fabry–Perot cavity (28~pm FWHM), temporally gated to 800~ps, and spatially filtered via single-mode fiber coupling. A combination of spectral, spatial, and temporal filtering minimized background noise. Real-time key extraction was achieved via integrated bidirectional laser communication, with synchronization and error correction performed using LDPC coding. The transmitter beam had a divergence of $\sim$20~$\mu$rad, producing a footprint that matched the 1.2~m receiver telescope. Fig.~1 depicts the full system configuration, including transmitter and receiver optics, tracking systems, filtering components, and real-time processing modules.

In 2020, during May 31–June 13, the system maintained continuous QKD operation under various challenging conditions. Despite intermittent weather, the setup generated over 42.7 Mbits of secure keys, with an average final key rate of 495 bps. Hourly link efficiency ranged from –43.3 to –34.7 dB, and the quantum bit error rate (QBER) remained consistently low (0.87\%–2.16\%), even under direct sunlight. These results were attributed to effective noise suppression and precise filtering, which kept dark counts below 250 cps. Beam quality assessment revealed a far-field divergence of $\sim$60~$\mu$rad, and atmospheric turbulence was characterized using the $R_{0}$ coherence length. Performance closely matched that of satellite-to-ground QKD systems, confirming the feasibility of real-time key generation under realistic Free-space conditions. Fig.~2 summarizes performance metrics over a 24-hour period, showing hourly trends in link efficiency, atmospheric coherence length, QBER, and final secure key rate.

\begin{figure*}[htbp]
    \centering
    \scalebox{0.5}{\includegraphics{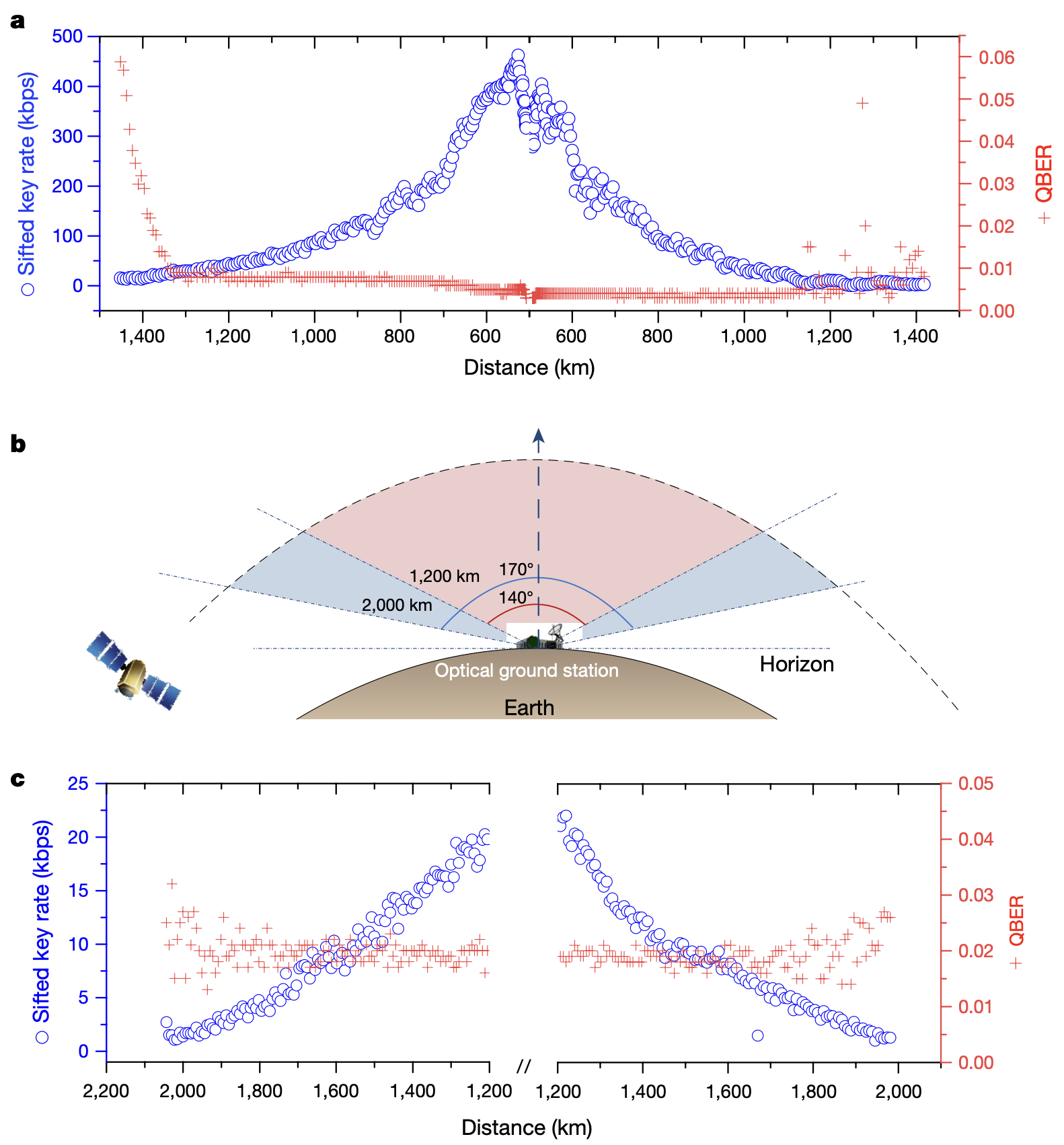}}
     \caption{Performance of high-speed satellite-to-ground QKD. (a) Sifted key rate (blue circles; left axis) and observed quantum-bit error rate (QBER; red plusses; right axis) as a function of slant range from the satellite to the Nanshan ground station. (b) Illustration of the coverage angle for high-speed satellite–ground QKD: the coverage angle (communication distance) is extended from about $140^{\circ}$ ($\sim\!1{,}200\,\mathrm{km}$; red) to about $170^{\circ}$ ($\sim\!2{,}000\,\mathrm{km}$; blue). (c) Long-distance satellite–ground QKD test, showing sifted key rate (blue circles; left axis) and observed QBER (red plusses; right axis) at distances exceeding $1{,}200\,\mathrm{km}$\cite{chen2021integrated}}
    \label{Fig_Nat2}
\end{figure*}

\begin{figure*}[htbp]
    \centering
    \scalebox{0.4}{\includegraphics{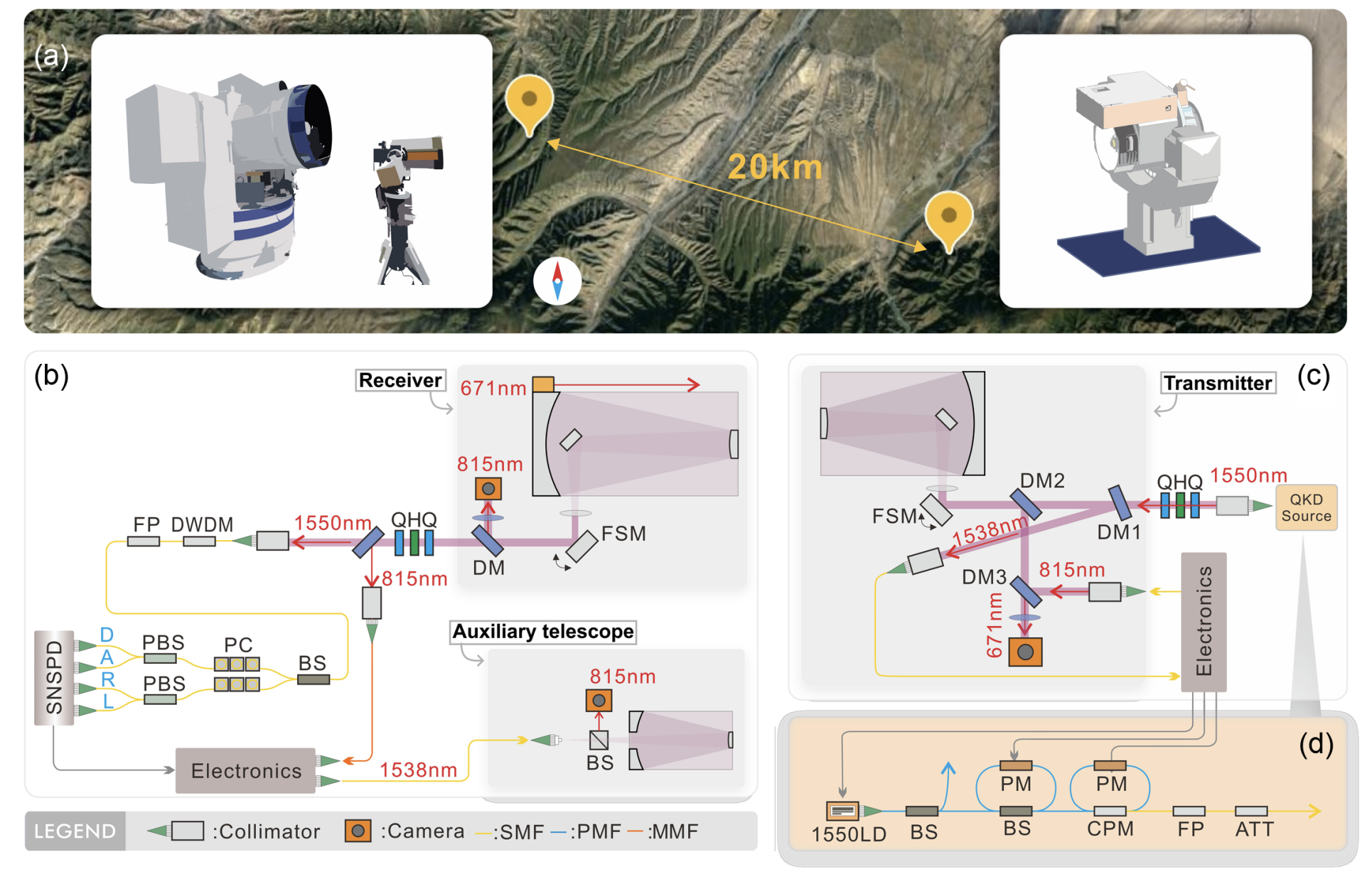}}
     \caption{All-day real-time Free-space QKD over a $20\,\mathrm{km}$ terrestrial link. (a) Field layout of the $20\,\mathrm{km}$ channel operated continuously through day and night. (b) QKD receiver. (c) QKD transmitter. (d) High-speed robust QKD light source. The transmitter and receiver first acquire and maintain the terrestrial optical link via acquiring–pointing–tracking (APT) systems (see Supplement~1) using an $815\,\mathrm{nm}$ downlink beacon and a $671\,\mathrm{nm}$ uplink beacon. Quantum signals at $1550\,\mathrm{nm}$ are prepared by the high-speed robust source, transmitted from the sender, and collected/detected at the receiver. Real-time key distillation uses an $815\,\mathrm{nm}$ downlink communication laser and a $1538\,\mathrm{nm}$ uplink communication laser. In the transmitter, DM1 reflects $1538\,\mathrm{nm}$ and transmits $1550\,\mathrm{nm}$; DM2 reflects $650$–$850\,\mathrm{nm}$ and transmits $1538$–$1550\,\mathrm{nm}$; DM3 reflects $671\,\mathrm{nm}$ and transmits $815\,\mathrm{nm}$. In the receiver, the auxiliary telescope (panel b, bottom right) is used to transmit $1538\,\mathrm{nm}$. Abbreviations: LD, laser diode; BS, (fiber) beam splitter; PM, phase modulator; CPM, customized polarization module; FP, Fabry–Perot filter; ATT, attenuator; SMF, single-mode fiber; PMF, polarization-maintaining fiber; MMF, multimode fiber; M, mirror; CAM, camera; FSM, fast-steering mirror; DM, dichromatic mirror; Q, quarter-wave plate; H, half-wave plate; DWDM, $100\,\mathrm{GHz}$ dense wavelength-division multiplexer (coarse filtering); PC, fiber polarization compensator; PBS, fiber polarization beam splitter; SNSPD, superconducting-nanowire single-photon detector. Adapted from W.-Q.~Cai \emph{et~al}\cite{cai2024free}.}
    \label{Fig_Optica_1}
\end{figure*}

The integration of high-speed modulation, robust filtering, and real-time classical communication resolved longstanding limitations associated with daylight operation and delayed key processing. The experimental conditions closely mirrored those expected in future low-Earth orbit (LEO) satellite scenarios, confirming system viability. Looking ahead, enhancements such as adaptive optics, finer temporal filtering, and operation at visible wavelengths are expected to further improve system robustness and scalability, paving the way for global quantum communication via high-orbit satellite constellations.

\subsection{Micro-satellite-based real-time quantum key distribution}

Li et al. (2025) presented the first real-time satellite-to-ground QKD using a lightweight micro-satellite and portable optical ground stations (OGSs). This work marked a major step toward a scalable quantum satellite constellation, enabling secure global communication via a cost-effective, rapidly deployable infrastructure. In a single satellite pass, they achieved up to 1.07 million secure bits and demonstrated encrypted intercontinental communication between China and South Africa.

\begin{figure*}[htbp]
    \centering
    \scalebox{0.45}{\includegraphics{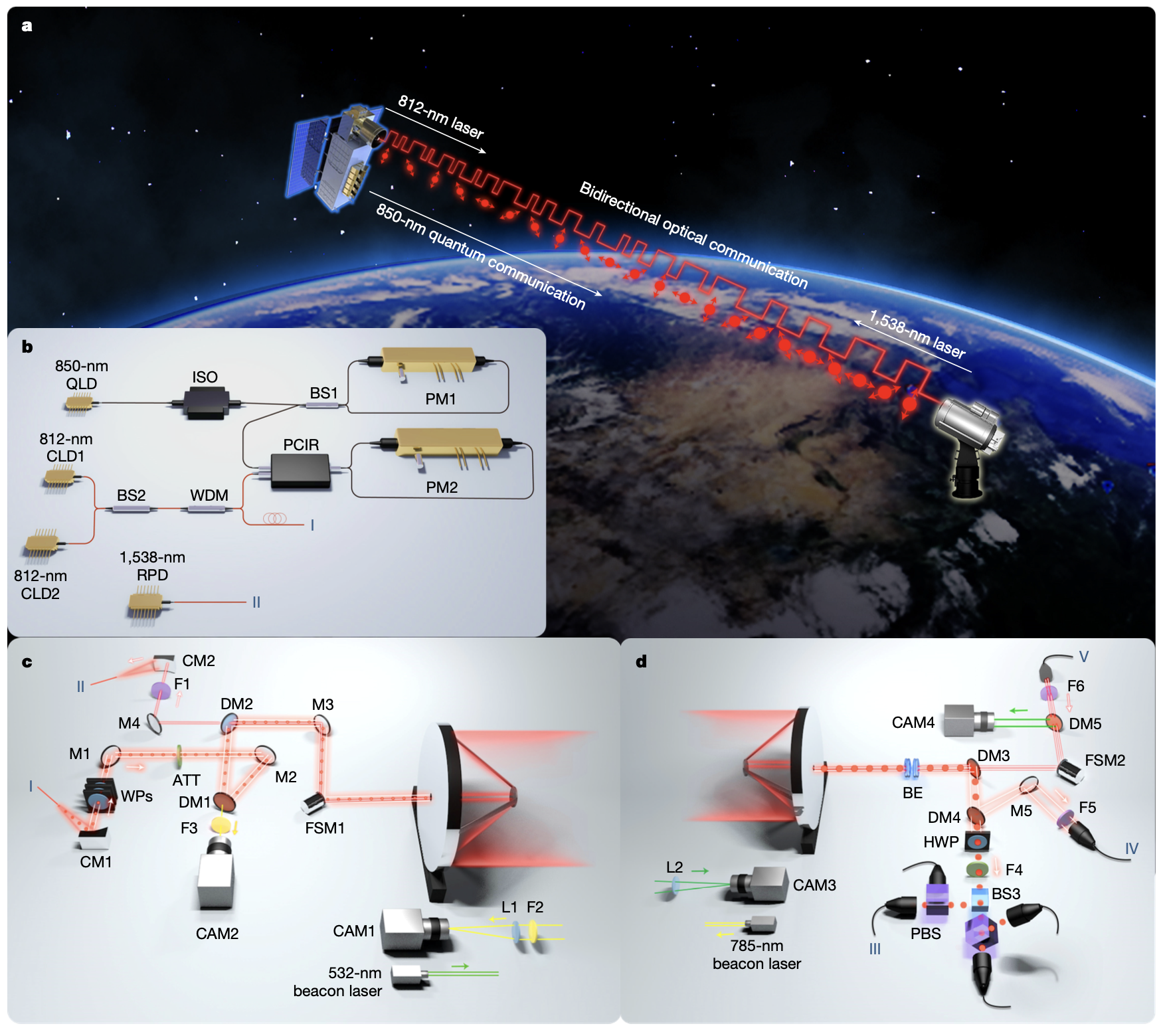}}
     \caption{Experimental set-up for satellite-to-ground QKD. (a) Overview of the satellite–ground experiment. In addition to the downlink $850\,\mathrm{nm}$ quantum photons, the satellite and the optical ground station (OGS) provide bidirectional optical communication—an $812\,\mathrm{nm}$ downlink laser and a $1{,}538\,\mathrm{nm}$ uplink laser—enabling real-time key distillation and secure communication. Map data: Google Earth, Data SIO, NOAA, US Navy, NGA, GEBCO, Landsat/Copernicus, IBCAO. (b) QKD light source based on a single laser diode with external modulation. Abbreviations: QLD, quantum laser diode; CLD, communication laser diode; ISO, isolator; BS, beam splitter; PM, phase modulator; PCIR, polarization module; RPD, receiving avalanche photodiode; WDM, wavelength-division multiplexer. (c) Satellite optical design. Label~I marks the output single-mode fiber (SMF) carrying the $812\,\mathrm{nm}$ communication beam and the $850\,\mathrm{nm}$ quantum signal; Label~II marks the receiving multimode fiber (MMF) for the $1{,}538\,\mathrm{nm}$ uplink. CAM1 is the capture camera; CAM2 is the fine-tracking camera. Additional optics: CM, concave mirror; WP, wave plate; M, mirror; ATT, attenuator; DM, dichroic mirror; F, filter; L, lens; FSM, fast-steering mirror. (d) Portable OGS. Labels~III,~IV, and~V mark, respectively, the receiving MMFs for the $850\,\mathrm{nm}$ quantum photons and the $812\,\mathrm{nm}$ downlink, and the transmitting SMF for the $1{,}538\,\mathrm{nm}$ uplink. BE, beam expander; PBS, polarization beam splitter. Adapted from Y.~Li \emph{et~al}\cite{li2025microsatellite}.}
    \label{Nature25a}
\end{figure*}

The micro-satellite Jinan-1, launched into a 500-km Sun-synchronous orbit, carried a 22.7 kg QKD payload and a 200-mm aperture telescope. Its quantum light source comprised a single 850 nm laser diode modulated via a Sagnac-interferometer-based scheme to implement a 625 MHz BB84 protocol with decoy states, supporting high repetition rates and resilience against side-channel attacks. Polarization compensation was performed using motorized wave plates onboard and at the ground station. 

\begin{figure*}[htbp]
    \centering
    \scalebox{0.5}{\includegraphics{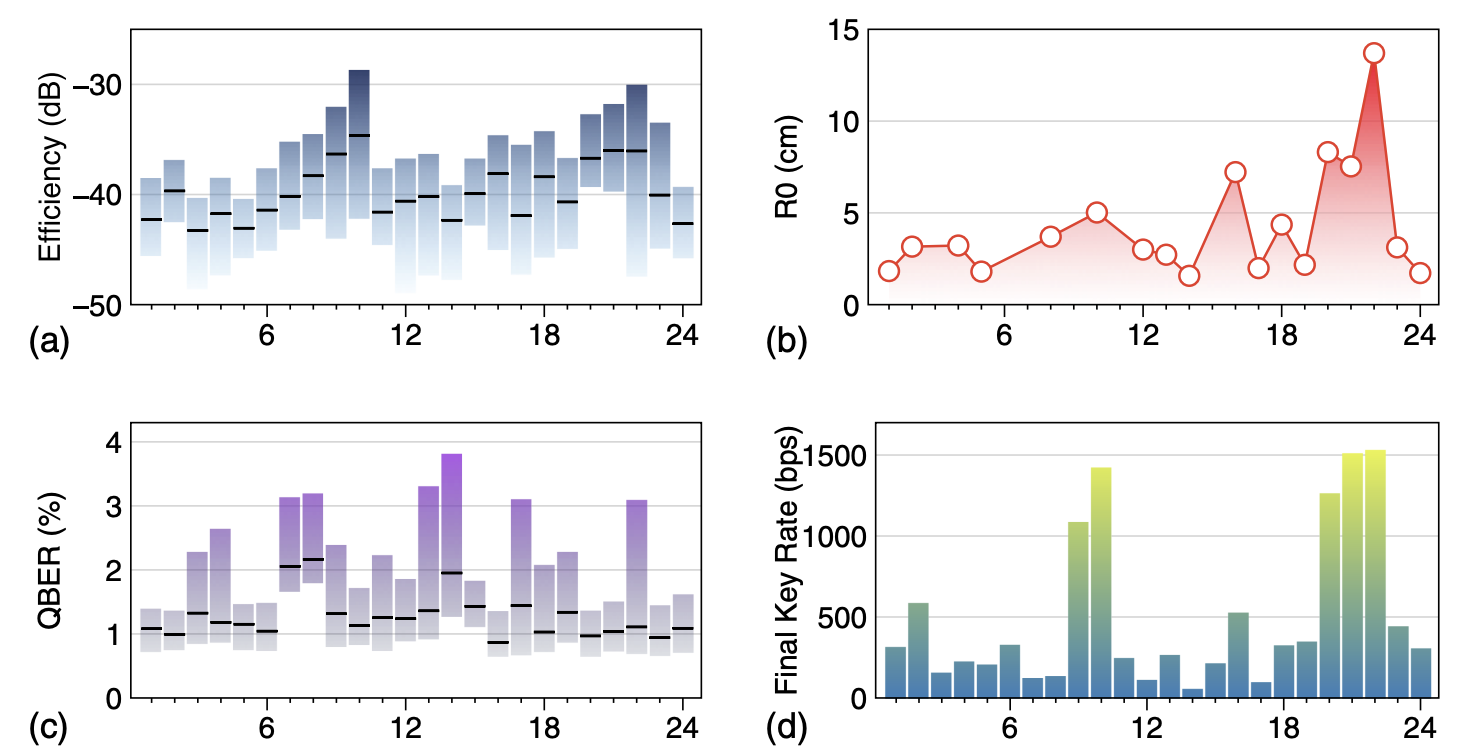}}
     \caption{Experimental results of Free-space QKD. (a) Link efficiency. (b) $R_{0}$. (c) Quantum-bit error rate (QBER). (d) Final key rate. Data were collected on different days between May~31 and June~13. The $R_{0}$ tests were performed without fine-tracking in the minutes preceding the QKD runs, whereas the efficiency, QBER, and final key-rate points are averages over runs of approximately $1\,\mathrm{h}$. Adapted from W.-Q.~Cai \emph{et~al}\cite{cai2024free}.}
    \label{Fig_Optica_2}
\end{figure*}

Fig. 1 outlines the system layout. Compact OGSs (~100 kg each) were deployed in both urban locations and remote areas. These ground stations featured 280-mm Cassegrain telescopes, polarization analysis modules, narrow bandpass filters, and low-noise silicon avalanche photodiode detectors (<800 cps dark counts). A two-stage acquisition, pointing, and tracking (APT) system enabled microradian-level precision. Fig. 3 shows the design of the satellite and OGS. Real-time bidirectional laser communication ensured precise timing synchronization (~100 ps) and data exchange at ~104 Mbps, enabling immediate key distillation during each satellite pass.

The system successfully performed real-time QKD across 20 orbits, linking ground stations in China and South Africa. During a pass over Jinan on 25 September 2022, secure key exchange lasted approximately 6 minutes. QKD commenced after the satellite achieved stable APT and the elevation exceeded 10–15°. The slant range varied from ~2,000 km to 500 km. Fig. 4 presents key performance metrics: photon detection rates rose with decreasing distance (4b), and QBER remained between 0.76$\%$ and 1.79$\%$ (4c), resulting in 406,784 secure bits distilled in real time (4a). Performance across all sites remained consistent, with average QBERs below 2.5$\%$. The best result—1.07 million secure bits in a single pass—was obtained at Stellenbosch, South Africa. Table 1 summarizes system performance across stations, confirming effective operation in diverse lighting and geographic conditions. Accurate polarization compensation and resilience to ambient light further validated system robustness. The use of low-density parity-check (LDPC) codes enabled efficient error correction with minimal overhead. A six-step real-time distillation protocol—including photon detection, synchronization, basis reconciliation, error correction, and privacy amplification—allowed final secure keys to be extracted on-the-fly during each pass. An intercontinental QKD demonstration was also conducted between China and South Africa (12,900 km apart), using the satellite as a "space postman" to relay secure keys. The resulting keys were applied to one-time pad encryption of image files, achieving practical, near real-time secure communication with a total delay of ~1.5 hours.

\begin{figure*}[htbp]
    \centering
    \scalebox{0.5}{\includegraphics{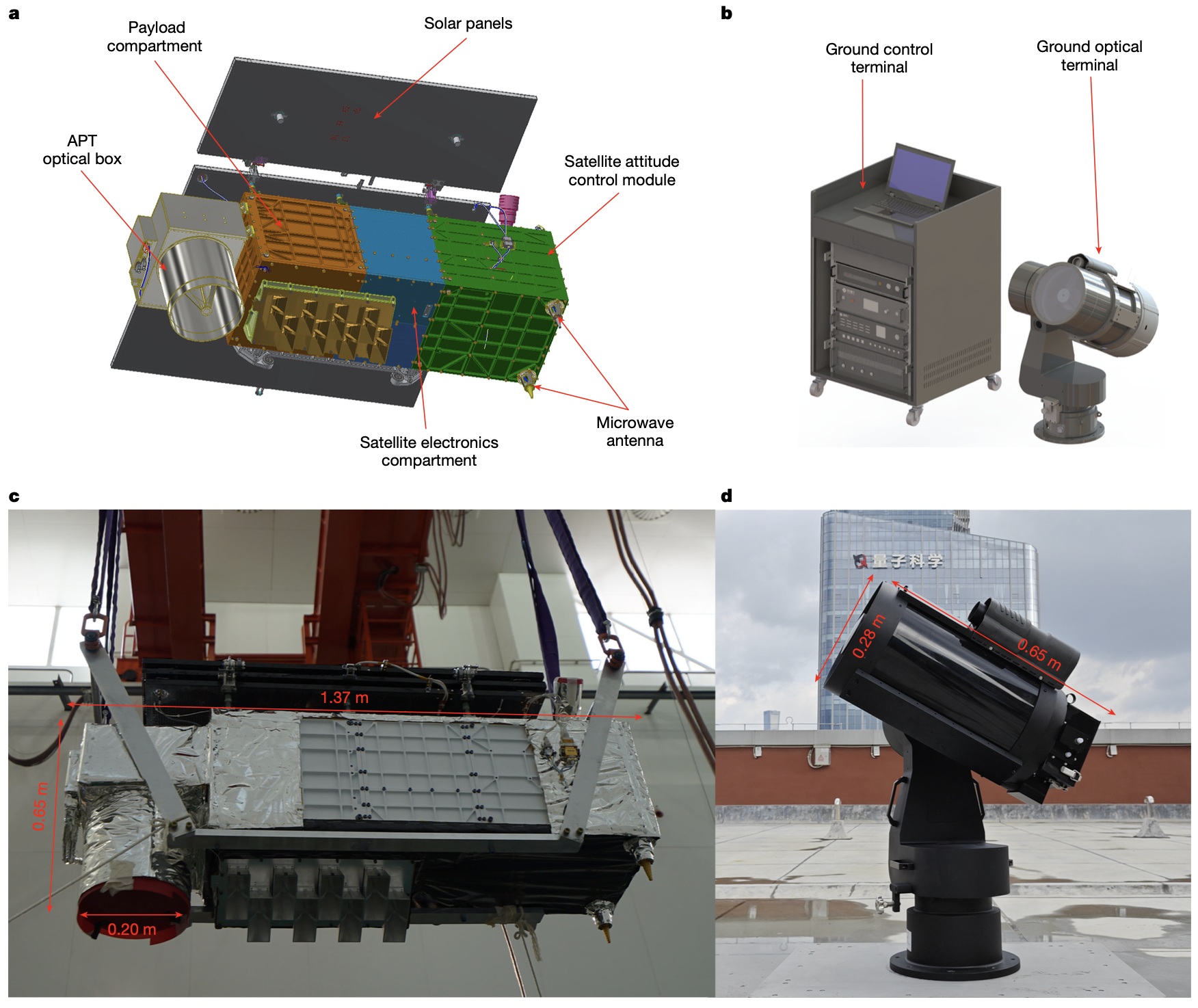}}
     \caption{Microsatellite and portable optical ground station (OGS). (a) The microsatellite comprises an acquisition–pointing–tracking (APT) optical box, payload compartment, satellite electronics compartment, attitude-control module, microwave antenna, and solar panels. (b) The OGS consists of a control terminal and an optical terminal. (c) Photograph of the microsatellite prior to rocket integration; the launch-state envelope is approximately $1.37\,\mathrm{m}\times 0.49\,\mathrm{m}\times 0.65\,\mathrm{m}$, with a telescope aperture of $0.2\,\mathrm{m}$. (d) Photograph of the portable OGS deployed in urban Jinan; the main telescope envelope is approximately $0.65\,\mathrm{m}\times 0.28\,\mathrm{m}\times 0.28\,\mathrm{m}$. Adapted from Y.~Li \emph{et~al}\cite{li2025microsatellite}}
    \label{Nature25b}
\end{figure*}
This work demonstrated a scalable and efficient architecture for space-based QKD using microsatellites and portable ground stations. By reducing the payload and ground system mass by over an order of magnitude compared to earlier works (e.g., Micius), the design offered rapid deployment, reduced cost, and high performance. The integration of real-time classical communication and fast key distillation eliminated delays associated with earlier satellite QKD experiments. The system consistently achieved high key rates (up to 1.07 Mbits per pass), low QBER (<2.5$\%$), and sub-microradian tracking precision—even under real-world atmospheric conditions. Its flexibility and portability laid the groundwork for scalable satellite constellations. Looking ahead, future improvements may include chip-scale light sources, daylight operation, and deployment in diverse orbits to expand coverage. These advances could support the development of a global quantum internet, enabling ultra-secure communication, long-distance quantum entanglement distribution, and secure access for users worldwide.

\begin{figure*}[htbp]
    \centering
    \scalebox{0.5}{\includegraphics{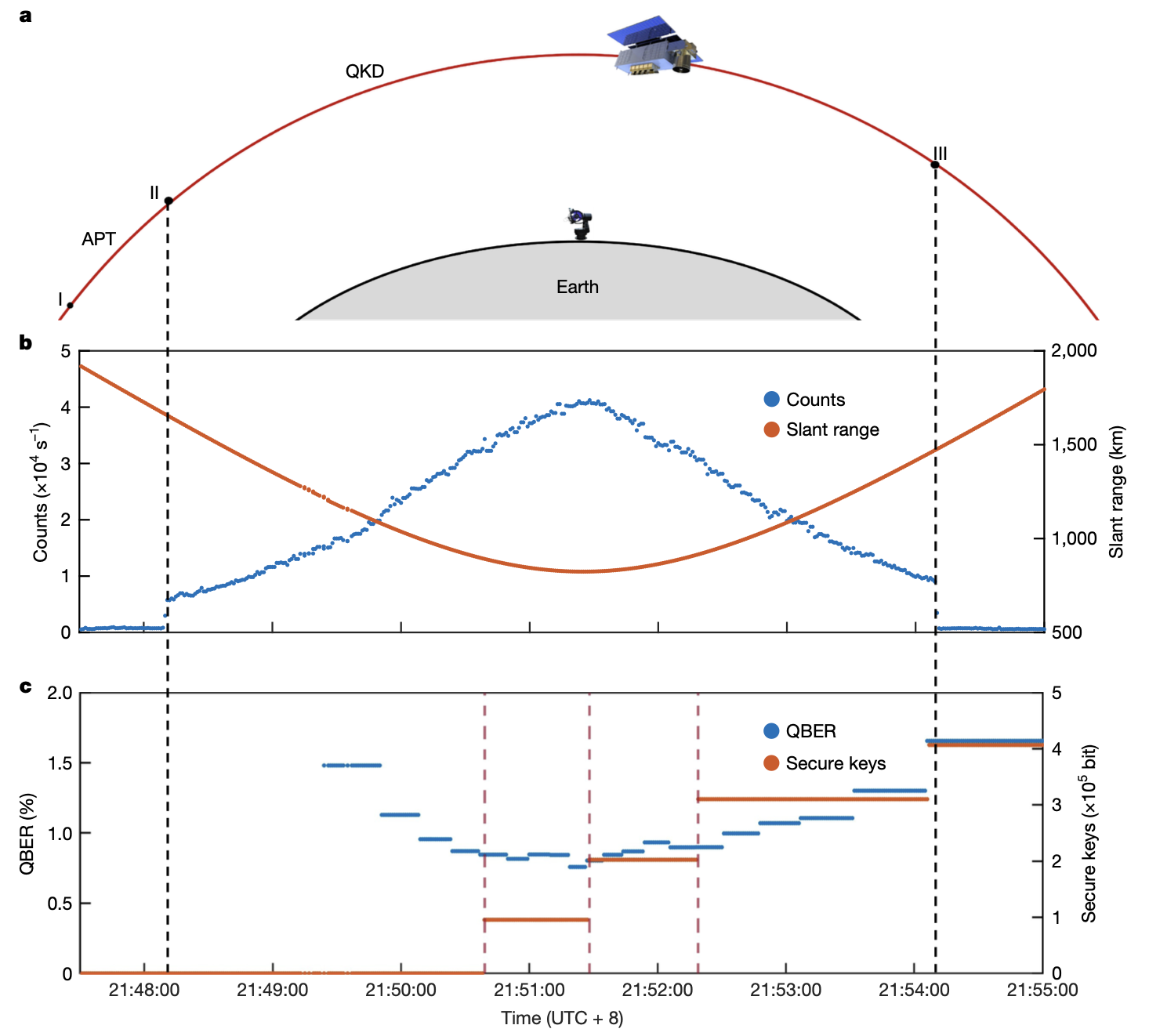}}
     \caption{Experimental set-up for satellite-to-ground QKD. (a) Overview of the satellite–ground experiment. In addition to the downlink $850\,\mathrm{nm}$ quantum photons, the satellite and the optical ground station (OGS) support bidirectional optical communication—an $812\,\mathrm{nm}$ downlink laser and a $1{,}538\,\mathrm{nm}$ uplink laser—enabling real-time key distillation and secure communication. Map data: Google Earth, Data SIO, NOAA, US Navy, NGA, GEBCO, Landsat/Copernicus, IBCAO. (b) QKD light source based on a single laser diode with external modulation. Abbreviations: QLD, quantum laser diode; CLD, communication laser diode; ISO, isolator; BS, beam splitter; PM, phase modulator; PCIR, polarization module; RPD, receiving avalanche photodiode; WDM, wavelength-division multiplexer. (c) Satellite optical design. Labels~I and~II denote, respectively, the output single-mode fiber (SMF) carrying the $812\,\mathrm{nm}$ communication beam and the $850\,\mathrm{nm}$ quantum signal, and the receiving multimode fiber (MMF) for the $1{,}538\,\mathrm{nm}$ uplink. CAM1 is the capture camera; CAM2 is the fine-tracking camera. Additional optics: CM, concave mirror; WP, wave plate; M, mirror; ATT, attenuator; DM, dichroic mirror; F, filter; L, lens; FSM, fast-steering mirror. (d) Portable OGS. Labels~III,~IV, and~V denote, respectively, the receiving MMFs for the $850\,\mathrm{nm}$ quantum photons, the receiving MMF for the $812\,\mathrm{nm}$ downlink laser, and the transmitting SMF for the $1{,}538\,\mathrm{nm}$ uplink. BE, beam expander; PBS, polarization beam splitter. Adapted from Y.~Li \emph{et~al}\cite{li2025microsatellite}.}
    \label{Nature25c}
\end{figure*}

\section{Atmospheric Turbulence and its Correction in Free-space QKD}
\label{sec:turbulence}

\subsection{Physical Origin and Metrics}
Atmospheric turbulence stems from stochastic fluctuations of temperature, pressure, and humidity, which imprint random refractive-index inhomogeneities along an optical path. The classical structure-function parameter $C_n^{2}(z)$ and the dimensionless \emph{Rytov variance} $\sigma_{\mathrm{Ry}}^{2}$ remain the workhorses for quantifying strength. For plane waves, the variance is defined as $\sigma_{\mathrm{Ry}}^{2}=1.23\,C_n^{2}k^{7/6}z^{11/6}$, with $k=2\pi/\lambda$ and path-length $z$ \cite{Ghalaii2022}. It is important to note that for spherical waves (approximating satellite downlinks), the Rytov variance scales as $0.4\times$ the plane wave value due to beam divergence \cite{Ghalaii2022}.

Weak, moderate, and strong regimes are conventionally separated by $\sigma_{\mathrm{Ry}}^{2}\lesssim0.3$, $\sim1$, and $\gtrsim1$, respectively. Satellite down-links routinely migrate into the latter two, especially near the horizon where refraction elongates the turbulent slab \cite{Sidhu2021}. Crucially, while the Rytov approximation for intensity fluctuations breaks down in the strong focusing regime, recent wave-optics simulations have confirmed that the classical analytical formulation of the Fried parameter ($R_{0}$) remains accurate well into the strong scattering regime (up to $\sigma_{\mathrm{Ry}}^{2} \approx 26.7$) \cite{Zhan2020}. This validates the continued use of $R_{0}$ as a fundamental design parameter for satellite links without requiring complex corrections for strong scattering.

The refractive index structure parameter $C_n^2$ is defined as \cite{tatarski2016wave}:
\begin{equation}
C_n^2 = \frac{\langle [n(\vec{x}) - n(\vec{x} + \vec{r})]^2 \rangle}{r^{2/3}},
\end{equation}
where $n$ is the refractive index, $\vec{x}$ denotes position along the propagation path, and $\vec{r}$ is the separation vector. The parameter $C_n^2$ typically ranges from $10^{-17}$ m$^{-2/3}$ for weak turbulence to $10^{-13}$ m$^{-2/3}$ for strong turbulence conditions.

\subsection{Turbulence Statistics and Channel Modeling}
A quantum channel through turbulence is \emph{fading}: the instantaneous transmissivity~$\eta$ follows a probability-density~$P(\eta)$. For weak turbulence ($\sigma_{\mathrm{Ry}}^{2}\lesssim1$), a log-normal model is adequate \cite{Erven2012}:
\begin{equation}
\mathcal{P}(\eta_{\text{atm}}) = \frac{1}{\sqrt{2\pi}\sigma\eta_{\text{atm}}} \exp\left[-\frac{1}{2}\left(\frac{\ln\eta_{\text{atm}} + \overline{\theta}}{\sigma}\right)^2\right].
\end{equation}
Beyond that, hybrid log-normal or elliptic-beam distributions are superior and still analytically tractable for security proofs \cite{Ghalaii2022}. These models predict that beam-spread (short-term broadening), rather than beam wander, dominates key-rate degradation once $\sigma_{\mathrm{Ry}}^{2}\!\gtrsim\!1$ \cite{Ghalaii2022}.

The impact of turbulence on an optical beam can be modeled using phase screens with aberrations described by Zernike polynomials:
\begin{equation}
\xi(\rho,\phi) = \sum_{n,m} c_{n,m} Z_{n,m}(\rho,\phi),
\end{equation}
where $Z_{n,m}$ are Zernike modes and $c_{n,m}$ are stochastic coefficients representing the instantaneous amplitude of each aberration. The statistical variance of these coefficients, defined as $\sigma_{n,m}^2 = \langle |c_{n,m}|^2 \rangle$, scales with the turbulence strength via the Fried parameter $R_{0}$. According to standard turbulence theory \cite{Noll1976}:
\begin{equation}
\sigma_{n,m}^2 = \gamma_{n,m} \left(\frac{D}{R_{0}}\right)^{5/3}, \quad \text{with} \quad R_{0} \approx 1.68 (C_n^2 L k^2)^{-3/5}.
\end{equation}

This expression for the Fried parameter assumes the plane-wave approximation.

\subsection{Impact on QKD Performance}
\paragraph{Fluctuating loss versus static-loss approximations.} Using a realistic decoy-state BB84 simulator, \citeauthor{Erven2012} showed the secure-key penalty caused by log-normal fading is negligible until turbulence becomes \emph{extremely} strong, validating the common static-loss approximation for most metropolitan links \cite{Erven2012}. However, the same work revealed opportunities: discarding blocks with poor signal-to-noise (\emph{SNR filtering}) can boost the final secret key by~$\sim25\%$ by rejecting high-QBER frames during deep fades \cite{Erven2012}.

\paragraph{Spatial-mode sensitivity.} Mode structure matters. A controlled laboratory comparison demonstrated that plane-wave (PW) encoding tolerates scintillation better than orbital-angular-momentum (OAM) modes beyond 1–2~km in daytime seeing, with PW modes maintaining three times lower crosstalk under identical turbulence conditions \cite{Mirhosseini2014}. This advantage stems from the definition of OAM being dependent on the optical axis; tip-tilt aberrations along the propagation axis cause mode-mixing in OAM, whereas one-dimensional PW modes are unaffected by wavefront tilts along the orthogonal direction \cite{Mirhosseini2014}. 

In quantum communications, turbulence-induced distortions introduce errors in high-dimensional encoding schemes. For OAM modes used in a $d$-dimensional 2-MUB protocol (generalizing BB84), the asymptotic secure key rate $R(d,e_q)$ under turbulence is defined by Eq. \ref{freq}
Security is compromised ($R < 0$) when the error rate exceeds a specific threshold (e.g., $\approx 11\%$ for $d=2$, increasing for higher $d$); for OAM links in the kilometer range, this breakdown typically occurs when turbulence strength exceeds $C_n^2 \gtrsim 10^{-15}$ m$^{-2/3}$.
Closely related wave–propagation phenomena have also been explored in classical analog systems. In particular, surface-gravity water waves have been shown to provide a controllable platform for studying dispersion, wave-packet spreading, interference, and effective potential landscapes under conditions mathematically analogous to paraxial optical propagation. Such hydrodynamic analogs offer intuitive insight into wave evolution, decoherence, and mode coupling in complex media, complementing numerical phase-screen and wave-optics models used for Free-space quantum channels \cite{GGR1,GGR2,GGR3,GGR4}.

\subsection{Passive Mitigation Strategies}
\paragraph{Aperture averaging and receiver diversity.} For Gaussian beams with long-term waist $w_{\mathrm{lt}}\gg a_R$ (receiver radius), averaging over multiple sub-apertures suppresses scintillation. A three-branch coherent-detection receiver utilizing optical combining achieved a $4$–$6$\,dB improvement in the $Q$-factor without active optics, demonstrating a practical path for satellite ground stations \cite{Yuan2020}.

\paragraph{Coherent noise suppression.} 
In Continuous Variable (CV) QKD, the local oscillator (LO) acts as a powerful spatial and spectral filter. By interfering the signal with the LO, only the mode matching the LO is detected. This effectively narrows the spectral filter bandwidth to the pulse bandwidth (e.g., $\Delta\lambda \approx 0.1$ pm), suppressing background noise by orders of magnitude compared to direct detection filters ($\sim 1$ nm). This mechanism renders daylight operation feasible even under high solar background \cite{Pirandola2021}.

\begin{figure}[htbp]
    \centering
    \scalebox{0.35}{\includegraphics{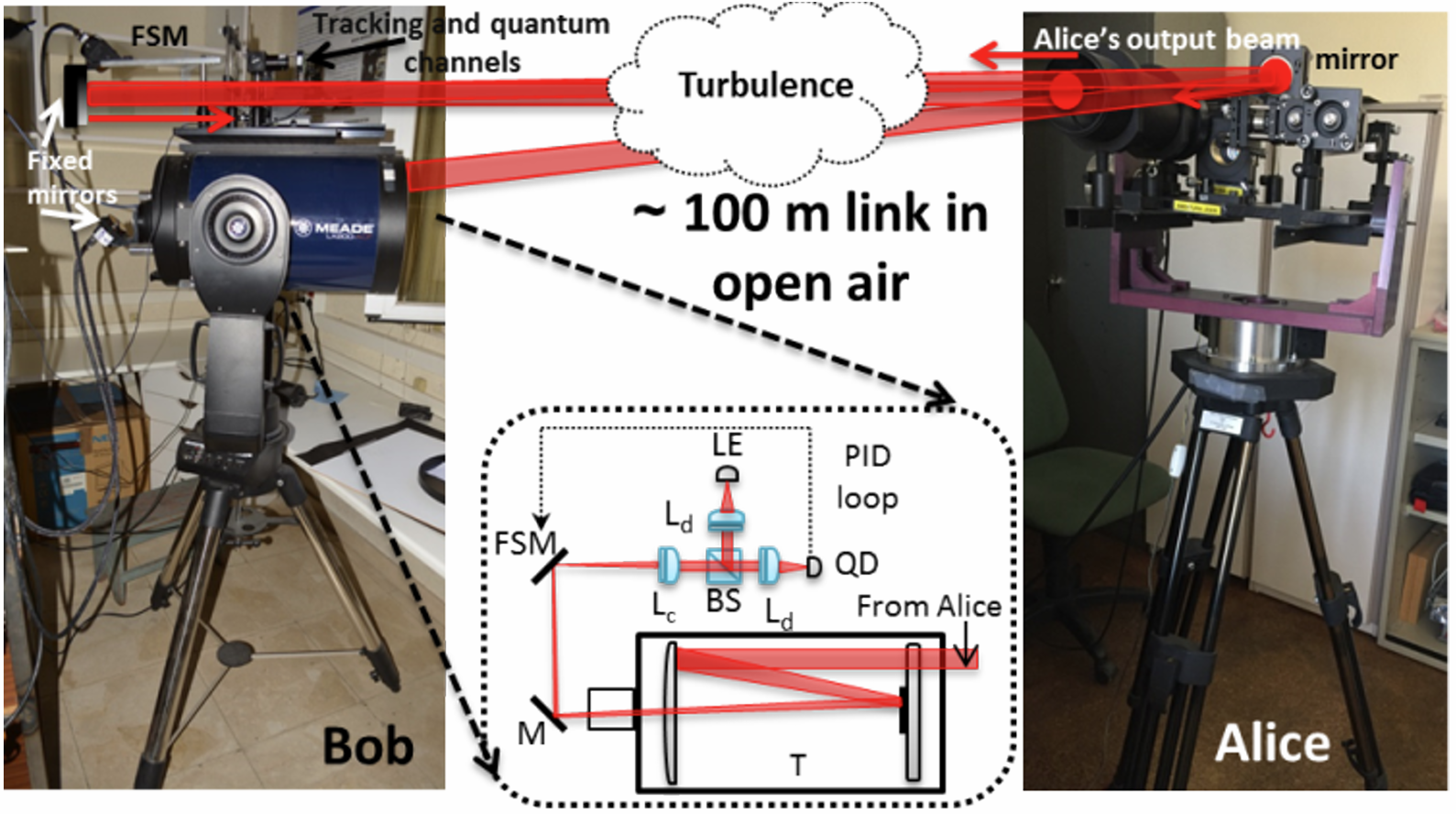}}
    \caption{Picture of the receiver and sender of the closed-loop correcting system at 100 m. T is a Schmidt-Cassegrain 25 cm-diameter telescope, M is a fixed mirror; FSM is a fast steering mirror; Lc and Ld are achromatic doublet lenses; BS is a 50/50 beamsplitter; LE and QD are lateral and quadrant position sensitive detectors; and PID is a proportional-integral-derivative loop \cite{Fernandez2018}.}
    \label{Fig_fig_turbo1}
\end{figure}

\subsection{Advanced Receiver Architectures}
To further mitigate fading and thermal noise in Discrete Variable (DV) systems, the Conditional Dynamics Kennedy (CD-Kennedy) receiver has been proposed \cite{Yuan2020}. Unlike static receivers, the CD-Kennedy receiver uses pilot bits to estimate the instantaneous turbulence fading $\eta_{eq}$. It dynamically adjusts the local oscillator amplitude to destructively interfere with the signal for the '0' bit state, effectively nulling the signal and reducing errors. This adaptive approach allows the receiver to surpass the standard quantum limit (SQL) in regimes of weak turbulence or low thermal noise, where static Kennedy and Type-II receivers typically fail \cite{Yuan2020}.

Photonic Integrated Circuits (PICs) offer a pathway to miniaturize these advanced architectures while enhancing robustness against vibrational and thermal drifts that plague bulk optics. Recent demonstrations have utilized silicon photonic chips containing meshes of tunable Mach-Zehnder Interferometers (MZIs) to act as coherent combiners \cite{Martinez2024}. While primarily characterized using classical signals to validate mode-mixing efficiency, the linear nature of these devices makes them directly applicable to quantum signals, where maximizing coupling into single-mode fibers is the critical challenge. In these devices, a 2D optical antenna array couples the distorted Free-space beam into waveguides, where the MZI mesh—controlled by local feedback loops—coherently sums the signals. This effectively replicates the function of a bulk aperture-averaging receiver but with significantly higher spatial resolution and mixing efficiency \cite{Martinez2024}.

For satellite downlinks where scintillation is severe, scalable PIC architectures such as 32-input coherent combiners have been developed \cite{DeMarinis2025}. These combiners can be paired with Multi-Plane Light Conversion (MPLC) devices, which spatially demultiplex the distorted wavefront into orthogonal modes (e.g., Hermite-Gaussian modes) before coherent combination on-chip \cite{Billault2021}. This "spatial demultiplexing" approach allows for the recovery of power that would otherwise be lost to mode-coupling errors in single-mode fiber coupling \cite{Ran2024}. Complementing PICs, dielectric metasurfaces are also emerging as ultra-compact interfaces for Free-space to fiber coupling, offering dynamic wavefront shaping with a footprint negligible compared to traditional adaptive optics mirrors \cite{Aldecocea2025}.

\subsection{Active Wave-front Correction}
Tip–tilt compensation is the first correction stage. Fernandez et al. demonstrated the use of quadrant detectors (QDs) and fast steering mirrors (FSMs) to correct wavefront tilt caused by beam wander in metropolitan QKD links \cite{Fernandez2018}. QDs are preferred over Lateral Effect (LE) detectors in quantum receivers because their response at the focal plane is maximized and independent of the Signal-to-Noise Ratio (SNR), whereas LE detectors struggle with the low SNR typical of quantum signals \cite{Fernandez2018}.

A closed-loop fast-steering mirror steered by a QD reduced focal-plane wander by a factor of~9 on a $100$-m test-range (Fig.~\ref{Fig_fig_turbo1}), corresponding to \(\approx10\times\) solar-background suppression for daylight QKD \cite{Fernandez2018}. Scaling rules indicate that for transmitter/receiver diameters $(D_T,D_R)=(4,8)$\,cm, receiver-only correction suffices up to $\sim1$–2.5\,km depending on $C_n^{2}$; beyond that, pre-compensation at the transmitter becomes essential \cite{Fernandez2018}.

\subsection{Machine Learning for Channel Estimation and Correction}
Recent advancements suggest that machine learning (ML) and deep learning (DL) can significantly outperform classical control loops in turbulence mitigation, particularly for predicting temporal fluctuations and correcting severe phase distortions. Conventional AO relies on wavefront sensors (like Shack-Hartmann) that struggle under strong scintillation. Deep learning approaches, specifically Convolutional Neural Networks (CNNs), have been demonstrated to predict turbulent phase screens directly from intensity images, bypassing the need for complex wavefront reconstruction \cite{Wang2021}.

Beyond wavefront correction, ML is increasingly used for real-time channel estimation. In Continuous Variable (CV) QKD, neural networks have been applied to predict channel transmittance and excess noise using pilot tones \cite{Liang2022}. This "ML-assisted" approach allows for more accurate post-selection thresholds compared to static statistical models, effectively increasing the secure key rate by adapting to transient atmospheric conditions \cite{Liang2022}. Furthermore, predictive models based on time-series analysis can forecast channel transmittance milliseconds ahead, potentially reducing the latency penalty in adaptive-rate protocols \cite{Yi2024}.

\subsection{Security Under Strong Turbulence}
Moderate-to-strong turbulence forces finite-key analyses to include the full $P(\eta)$ and excess-noise terms. Ghalaii \& Pirandola (2022) derived ultimate quantum communication limits for moderate-to-strong turbulence, revising the PLOB bound to include turbulence-induced beam widening. 
The overall transmissivity is modeled as the product of four distinct loss mechanisms \cite{Ghalaii2022}:
\begin{equation}
\eta = \eta_{\text{lt}} \eta_{\text{eff}} \eta_{\text{cd}} \eta_{\text{atm}},
\end{equation}
where:
\begin{itemize}
    \item $\eta_{\text{lt}}$ is the \emph{long-term turbulence transmissivity}, which accounts for the beam widening and breaking caused by refractive index fluctuations beyond the diffraction limit;
    \item $\eta_{\text{atm}}$ represents \emph{atmospheric extinction} modeled by the Beer-Lambert law, capturing loss due to absorption and scattering by aerosols and molecules;
    \item $\eta_{\text{eff}}$ denotes the static \emph{receiver efficiency}, summarizing optical losses in the telescope setup and the finite quantum efficiency of the detectors;
    \item $\eta_{\text{cd}}$ accounts for the \emph{coherent detection efficiency}, specifically the mode-matching loss relevant to Local Local Oscillator (LLO) schemes where the locally generated reference pulse does not perfectly overlap spatially with the distorted signal beam.
\end{itemize} 
For $\sigma_{Ry}^2 > 1$, beam wandering becomes negligible compared to beam widening, simplifying security analysis to a stable channel with high loss \cite{Ghalaii2022}.

A composable CV-QKD proof shows positive key rates remain attainable at $\sigma_{\mathrm{Ry}}^{2}\!\sim\!10$ provided homodyne detectors exhibit $\eta_{\mathrm{det}}\!\gtrsim\!60\%$ and electronic noise $v_{\mathrm{el}}\!\lesssim\!0.01$\,SNU \cite{Ghalaii2022}. Key findings include:
\begin{itemize}
    \item Positive key rates are achievable even at 10 km in strong turbulence ($\sigma_{Ry}^2 \approx 38$ at night) with realistic receivers.
    \item Increasing receiver aperture ($a_R > 30$ cm) can compensate for high background noise at the cost of increased thermal photon counts.
    \item For satellite links at large zenith angles (e.g., mask angle $\theta_m = 80^\circ$), CV-QKD remains feasible up to 500 km altitude with block sizes $\sim 10^{10}$–$10^{12}$.
\end{itemize}

\subsection{Open Problems and Outlook}
\label{subsec:outlook}

Several challenges remain open for future investigation, particularly as Free-space quantum communications transition from laboratory demonstrations to real-world deployments. In the \emph{deep-strong turbulence regime} ($\sigma_{\mathrm{Ry}}^{2}\!\gg\!10$), empirical data are scarce, and systematic measurement campaigns are required to accurately characterize performance and validate theoretical models like the elliptic-beam distribution \cite{Ghalaii2022}. Beyond these fundamental characterization needs, several promising research directions have emerged:

\paragraph{Hybrid active-passive mitigation architectures.} 
While individual techniques like adaptive optics (AO) \cite{Fernandez2018} and signal-to-noise ratio (SNR) filtering \cite{Erven2012} have shown promise, their synergistic integration remains underexplored. A coherent architecture combining real-time wavefront correction with intelligent frame selection could yield additive improvements. For instance, AO could stabilize the beam centroid to maximize coupling, while SNR filtering would reject residual frames with elevated QBER due to uncorrected scintillation.

\paragraph{Standardized turbulence characterization.}
The field lacks standardized metrics for quantum channels. Classical parameters like $C_n^2$ and $R_{0}$ do not fully capture quantum state degradation. Developing quantum-specific tools—such as direct measurement of entanglement degradation—would enable more accurate performance predictions \cite{Erven2012}.

\paragraph{Fully-integrated photonic receivers.}
The transition from bulk optics to chip-scale receivers is a critical frontier for satellite payloads facing strict SWaP (Size, Weight, and Power) constraints. Future research must address the efficiency of Free-space-to-chip coupling under severe turbulence using metasurfaces and MPLC devices. Furthermore, the development of "self-adaptive" PICs that integrate turbulence sensing and correction on a single platform could eliminate the need for complex external AO loops \cite{Martinez2024, DeMarinis2025}.

\paragraph{AI-driven autonomous quantum links.}
Machine learning is poised to move beyond simple parameter estimation to full link autonomy. Future "cognitive" ground stations could use deep reinforcement learning to predict turbulence patterns and proactively adjust AO settings, minimizing latency errors. Additionally, end-to-end learning architectures could jointly optimize the transmitter modulation and receiver post-processing strategies in real-time, adapting the protocol (e.g., switching between DV and CV) to match the instantaneous atmospheric "weather" \cite{Liang2022}.

\begin{figure*}[htbp]
    \centering
    \scalebox{0.45}{\includegraphics{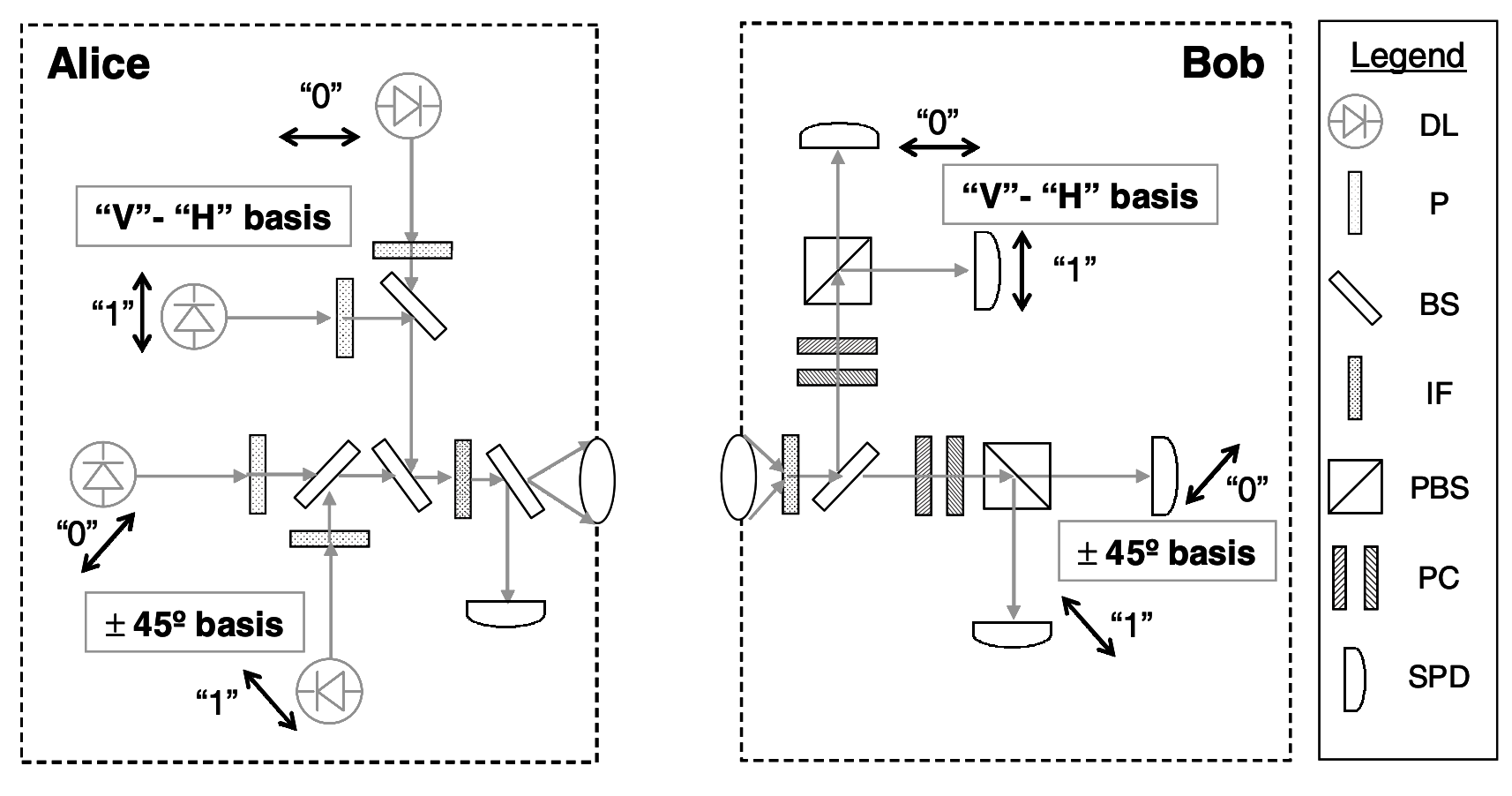}}
     \caption{Polarization-optics layout of the BB84 QKD transmitter (Alice) and receiver (Bob). Reproduced from Fig. 2 of Richard J. Hughes et al., New Journal of Physics 4, 43 (2002) \cite{hughes2002practical}. The outputs of Alice’s four data lasers (DL) are first heavily attenuated (mean photon number µ < 1), then linearly polarized to the BB84 states (double-headed arrows) by polarizers (P). After combination on non-polarizing beamsplitters (BS), the pulses traverse a spatial filter (not shown) to erase spatial-mode information, an interference filter (IF) to suppress spectral side-channels, and finally a 50:50 BS. Photons transmitted through this BS travel toward Bob, while those reflected are sampled by a single-photon detector (SPD) operating in a ~20 ns gate to monitor the launched µ-value; relative DL timings are matched to within the SPD timing jitter. At Bob, incoming pulses pass an IF and are randomly directed by the first BS into one of two analysis arms. In the reflected arm the rectilinear (H/V) basis is selected via a polarization controller (PC) and polarizing BS (PBS); a valid bit is registered when exactly one of the two SPDs at the PBS outputs fires inside the timing window. The transmitted arm performs an analogous measurement in the conjugate diagonal (±45°) basis. Multi-detection events (more than one SPD firing) are logged but excluded from key generation.}
    \label{Fig_QKD_Setup}
\end{figure*}

\subsection*{Practical checklist for system designers}
\begin{enumerate}
  \item Characterize $C_n^{2}(z)$ locally and convert to $\sigma_{\mathrm{Ry}}^{2}$, distinguishing between plane and spherical wave models for satellite links \cite{Zhan2020}.
  \item For short links ($<2$ km), prioritize tip–tilt correction; for longer links where scintillation dominates, aperture averaging often yields better returns \cite{Fernandez2018, Yuan2020}.
  \item Budget excess noise and fading variance in finite-key analysis; ignore beam-wander only when $w_{\mathrm{lt}}\!\gg\!w_{\mathrm{st}}$ (typically $\sigma_{\mathrm{Ry}}^{2} > 1$) \cite{Ghalaii2022}.
  \item Consider SNR filtering if background counts exceed 1–2\,\% of the signal to recover key from noisy data \cite{Erven2012}.
  \item Implement \emph{pilot-guided loss tracking} using energetic pulses to estimate instantaneous transmissivity $\eta$ in real-time, enabling accurate binning of signals for post-selection \cite{Pirandola2021}.
  \item For satellite links at low elevation angles, specifically account for path elongation and the increased effective turbulence strength \cite{Ghalaii2022}.
\end{enumerate}

\section{Standardization of Free-space QKD}

Standardization is essential for Free-space QKD as it ensures compatibility and reliability across different systems and implementations. By defining common protocols, hardware requirements, performance benchmarks, and test methods, standardization enables the seamless integration of QKD into existing communication networks. It also enhances interoperability between different vendors, facilitating widespread adoption and scalability, which is beneficial to open up a larger market. Moreover, standardized security frameworks help validate the robustness of Free-space QKD systems against potential threats, ensuring trust in QKD-based encryption for global cybersecurity applications.

\begin{table*}[htbp]
  \begin{center}
	\caption{Summary of work items on QKD under SDOs}
	\label{ITU_T_ETSI}
	\begin{tabular}{|p{66pt}|p{60pt}|p{125pt}|p{160pt}|}
		\hline
		Standards Group & Work item code & Title & Satellite-based QKD related content \\
	\hline
        \multirow{2}*{\makecell[l]{ITU-T \\FG-QIT4N}} & D2.2 & Quantum information technology for networks use cases: Quantum key distribution network & Discussing QKDN use cases including QKDN as Free-space satellite-ground or inter-satellite network, identifying architectural impacts and technical requirements.\\	
        \cline{2-4}
        ~ & D2.5 & Standardization outlook and technology maturity: Quantum key distribution network & presenting advancements in frontier research on QKD, with a focus on satellite-based QKD, highlighting it as an important aspect of standardizing key issues for QKDN.\\
        \hline
        \multirow{3}*{ITU-T SG13} & Y.3800 & Overview on networks supporting quantum key distribution & Providing conceptual layer structures of QKDN and defining QKDN capabilities. \\
	\cline{2-4}
        ~ & Y.3802 & Quantum Key Distribution networks - Functional architecture & Defining a functional architecture model of QKDN and specifying reference point to support satellite-based QKDN. \\
        \cline{2-4}
        ~ & TR.SQKDN & Standardization consideration of Satellite-based QKDN & Analysing the architecture and functional requirements of satellite-based QKDN \\
		\hline
        ITU-T SG17 & TR.SQKDN-SC & Security consideration for satellite-based quantum key distribution network & Providing research on potential security risks, security requirements, security measures of satellite-based QKDN. \\
        	\hline
        ETSI TC SES & DTS/SES-00469 & Satellite-Quantum Key Distribution (S-QKD) Satellite Systems \& Associated Optical Earth Stations (OES) & Specifying use cases, reference architectures, QKD protocols and technical/operational measures of satellite-based QKD systems. \\
        	\hline
	\end{tabular}
  \end{center}
\end{table*}

There are many international standard development organizations (SDOs) that devote significant efforts in defining Free-space QKD standards, the International Telecommunication Union (ITU) and European Telecommunications Standards Institute (ETSI) being the two major ones among them. 

ITU-T has been developing standards for quantum networks since 2018. It initially established a focus group on quantum information technology for networks (FG-QIT4N) to explore the evolution and applications of quantum information technologies in networking, with QKD being a key technology. The output deliverable of FG-QIT4N discusses the QKD network (QKDN) as a Free-space satellite-ground or inter-satellite network and outlines the expected architectural impact, technical requirements, as well as protocols, performance, and security requirements that will guide future standardization efforts. Based on foundational insights provided by FG-QIT4N, ITU-T study group (SG) 13 has been developing a series of standards for QKDN, including requirements, functional architecture, and service procedures, where supporting direct Free-space optical channels for quantum channel networking is identified as a capability for QKDN. In addition to these technical specifications, SG13 and SG17 also conduct technical reports to analyze the functional requirements and security aspects of the satellite-based QKDN. 

In ETSI, the standardization work on QKD is conducted in a dedicated Industry Specification Group (ISG). Different from ITU-T standardizing QKD at network-level, ISG QKD addresses device-level specifics such as components and interface. However, most of works in ISG QKD focuses on fiber optical network, specifications for Free-space QKD fall under the purview of ETSI's Technical Committee for Satellite Earth Stations and Systems (TC SES). TC SES has initiated a technical specification for Satellite-Quantum Key Distribution (S-QKD) Satellite Systems \& Associated Optical Earth Stations (OES) to specify the reference architectures, QKD protocols and technical/operational measures of satellite-based QKD systems. Details of the aforementioned work items in ITU-T and ETSI are summarized in Table~\ref{ITU_T_ETSI}.




\begin{table*}[htbp]
  \begin{center}
    \caption{Summary of key QKD experiments demonstrating increasing communication distance}
    \label{qkd_progress_summary}
    \begin{tabular}{|p{80pt}|p{70pt}|p{280pt}|}
      \hline
      \textbf{Protocol} & \textbf{Distance Achieved (km)} & \textbf{Main Significant Experimental Approach} \\
      \hline
      LED-based BB84 & 1.0 & Low-cost implementation using LEDs, passive optical components, and a single-photon detector, targeted for last-mile secure communications \\
      \hline
      CV-QKD (Discrete Signaling) & 24.2 & Post-selection technique, polarization multiplexing, and quantum state tomography to enhance secure key rate using continuous-variable measurements \\
      \hline
      Decoy-state BB84 & 90.0 & Advanced temporal filtering and CWDM multiplexing enabling coexistence with bidirectional 1 Gbps classical data over a single fiber \\
      \hline
      Entanglement-based QKD & 100.0 & Use of superconducting nanowire single-photon detectors (SNSPDs) and ultra-stable Mach-Zehnder interferometers to maintain entanglement fidelity over long distances \\
      \hline
      Measurement-Device-Independent QKD & 404.0 & Security enhancement via four-intensity decoy-state protocol, high-efficiency SNSPDs, and optimized parameter estimation over ultralow-loss fiber \\
      \hline
      Twin-Field QKD (SNS protocol) & 658.0 & Ultralong-distance communication enabled by phase-locked ultrastable lasers, heterodyne frequency calibration, and real-time phase compensation \\
      \hline
    \end{tabular}
  \end{center}
\end{table*}

\section{Challenges and Future Perspective}

Despite the remarkable progress in satellite-based quantum communication, several challenges remain. The primary technical challenge is the atmospheric turbulence that can induce decoherence and losses in the transmitted quantum signals. This turbulence is particularly significant during the final leg of the communication path between the satellite and the ground station. Various adaptive optics techniques and error-correction protocols are being developed to mitigate these effects and improve the fidelity of the received quantum states.

Another critical challenge is the development of quantum repeaters, which are necessary to extend the range of quantum communication beyond the limitations imposed by direct transmission. Quantum repeaters, which rely on entanglement swapping and quantum memory, are still in the experimental stage, but they hold the key to enabling long-distance, high-fidelity quantum communication. In parallel, research is ongoing to miniaturize quantum communication components and integrate them into smaller, more cost-effective satellite platforms, such as CubeSats, which could significantly reduce the cost and complexity of deploying a global quantum network.

The future of satellite-based quantum communication looks promising, with the potential for new protocols and technologies to emerge that will further enhance the security and efficiency of quantum communication systems. The ongoing development of quantum communication satellites, coupled with advancements in ground station technology and quantum repeater networks, will likely lead to the realization of a global quantum internet within the next decade.

\newpage
\begin{acknowledgments}

G.G.R. acknowledges support from the C. L. E. Moore Instructorship and from the MIT School of Science Research Innovation Seed Fund, supported by the Alfred P. Sloan Foundation.
\\
N. K. Kundu\ acknowledges the funding support from the National Quantum Mission of India, INSPIRE Faculty Fellowship (Reg. No.: IFA22-ENG 344), ANRF Prime Minister Early Career Research Grant (ANRF/ECRG/2024/000324/ENS), and the New Faculty Seed Grant from IIT Delhi. 

\end{acknowledgments}

\newpage

\bibliographystyle{unsrt}   
\bibliography{references} 

$\,$

$\,$

\end{document}